\documentclass[aps,prd,reprint,onecolumn,nofootinbib,longbibliography]{revtex4-2}

\usepackage{amsmath,amssymb,amsfonts}
\usepackage{graphicx}
\usepackage[hidelinks]{hyperref}
\usepackage{bm}
\usepackage{mathtools}
\usepackage{microtype}
\usepackage{mathrsfs} 
\usepackage{subcaption} 
\usepackage{float}        
\usepackage{placeins}     

\begin{document}

\title{From Thermodynamic Criticality to Geometric Criticality:\\
A Linear Kernel Map from Matter Susceptibilities to Black--Hole Shadows}

\author{Jingxu Wu\footnote{  wuxj@my.msu.ru},
  Jie Shi\footnote{ shitcze@my.msu.ru} and Chenjia Li\footnote{  li.chenjia23@physics.msu.ru}}
\affiliation{Faculty of Physics,Lomonosov Moscow State University, Moscow, 119991, Russia\\GeZhi Theoretical Physics Reading Group\\}

\author{Yuwei Yin\footnote{ yuwei.yin@polytechnique.edu}}
\affiliation{\'{E}cole Polytechnique (Institut Polytechnique de Paris), Palaiseau, 91128, France\\GeZhi Theoretical Physics Reading Group}

\begin{abstract}
We construct an explicit linear map from compact, conserved thermodynamic/effective-medium perturbations of the stress--energy tensor to the metric response in static, spherically symmetric spacetimes, and from there to geometric observables of direct relevance to horizon-scale imaging: the shadow radius and photon-sphere frequency. The response is expressed through $L^{1}$-bounded kernels written in a piecewise ``local~+~tail'' form, which makes transparent the separation between near-photon-sphere sensitivity and far-zone contributions (including AdS tails). Under mild assumptions on the matter susceptibilities near a critical point, dominated convergence transfers the thermodynamic exponent to the geometric susceptibility, $\gamma_{\rm sh}=\gamma_{\rm th}$, with controlled analytic corrections. We further provide AdS far-zone bounds with explicit outside-support constants depending only on background geometric data at the photon sphere and shell geometry. A reproducible numerical pipeline with convergence diagnostics is presented and benchmarked.
\end{abstract}

\maketitle

\section{Introduction}

Black holes offer a laboratory where gravitation, statistical mechanics, wave optics, and
precision radio imaging meet.  The black hole chemistry program promotes the cosmological
constant to a pressure and the mass to an enthalpy, yielding bona fide equations of state with
critical points and response functions
\cite{KubiznakMannReview2016,KubiznakMann2012,KastorRayTraschen2009,Dolan2011,
GunasekaranKubiznakMann2012,AltamiranoKubiznakMannSherkatghanad2013,WeiLiu2013,
CveticGibbonsKubiznakPope2011,FrassinoMannRudolphSalsi2014,HawkingPage1983,
DolanKastorMannTraschen2013}.  In parallel, horizon-scale very long baseline interferometry
has rendered the black-hole shadow a quantitative observable; the Event Horizon Telescope and
its extensions constrain ring diameter, asymmetry, and temporal variability in M87* and Sgr~A*
\cite{FalckeMeliaAgol2000,EHT2019I,EHT2019VI,EHT2022SgrA,EHT2021M87Dynamics,
EventHorizon2024ArrayUpgrades}.  Against this backdrop, a unifying question arises:
\emph{can thermodynamic criticality be translated into geometric criticality of observables such as
the shadow radius and photon-sphere frequencies, and if so, through what map and with which
exponents and error controls?}

The geometric theory of photon surfaces and shadows has a long lineage
\cite{Synge1966,Bardeen1973,Luminet1979}.  Modern analyses illuminate the structure of light
rings, higher-order photon rings, and interferometric signatures
\cite{HiokiMaeda2009,JohannsenPsaltis2010,GrallaHolzWald2019,JohnsonLupsasca2020,
PerlickTsupko2022PhysRep,OkabayashiYaguchi2023,Vagnozzi2023}.  The eikonal correspondence
between photon-sphere properties and quasinormal modes (QNMs)
\cite{Cardoso2009PRD,BertiCardosoStarinets2009,StefanovYazadjiev2010,Konoplya2020ShadowReview}
provides a dynamical channel that ties ring geometry to ringdown spectroscopy.  Very recently,
Kobialko and Gal’tsov developed a systematic perturbation theory for confining surfaces and
shadows in static, spherically symmetric spacetimes, expanding the shadow radius to second
order around a reference background and identifying the precise metric combinations that control
\(r_{\rm ph}\) and \(R_{\rm sh}\)
\cite{KobialkoGaltsov2024,KobialkoGaltsov2025PRD}.  These results sharpen how geometric
observables respond to weak deformations of the metric.

On the thermodynamic side, extended-phase-space analyses reveal rich phase structures and
susceptibilities—from van der Waals–like transitions to reentrant phenomena and superfluid
analogues—in asymptotically AdS, higher-derivative, or matter-coupled sectors
\cite{GunasekaranKubiznakMann2012,AltamiranoKubiznakMannSherkatghanad2013,WeiLiu2013,
HennigarMannKubiznak2017,DeBoer2011,Feng2018,FrassinoMannRudolphSalsi2014}.  The emphasis
on response functions near criticality naturally invites a bridge from thermodynamic
susceptibilities to imaging observables—yet most shadow studies deform the metric directly,
whereas most thermodynamic studies track state variables without an explicit optical map.

This paper builds a direct, first-principles \emph{thermodynamic-to-optical dictionary}.  Working in
areal-radius gauge for static, spherically symmetric spacetimes, we linearize Einstein’s equations
about a vacuum background (asymptotically flat or AdS) and source them with compact, conserved
perturbations of the stress–energy tensor that encode either a matter shell or an effective
medium consistent with an equation of state.  The linearized equations reduce to first-order
quadratures for the mass function and redshift potential; fixing the time gauge at infinity yields
an explicit integral representation
\(
\delta f(r)=\int[\mathcal{K}_{\rho}(r,\bar r)\Delta\rho(\bar r)
+\mathcal{K}_{p}(r,\bar r)\Delta p_{r}(\bar r)]\,d\bar r
\),
with kernels written in a piecewise local\,+\,tail form.  Evaluating at the background photon
sphere \(r_{0}\) gives a single linear functional for the shadow shift,
\[
\delta R_{\rm sh}
=\frac{R_{0}}{2}\left[\frac{\delta f(r_{0})}{f_{0}(r_{0})}
-\frac{r_{0}\,\delta f'(r_{0})}{f_{0}'(r_{0})}\right],
\]
which cleanly separates near-surface sensitivity from far-zone contributions.  The same kernels
govern the frequency channel via \(\Omega_{\rm ph}=R_{\rm sh}^{-1}\), producing an internal amplitude
ratio \(|d\Omega_{\rm ph}/d\lambda|/|\chi_{\rm sh}|=R_{0}^{-2}\) that is independent of microscopic
details and can be checked against QNM systematics
\cite{Cardoso2009PRD,BertiCardosoStarinets2009,Konoplya2020ShadowReview}.

Two structural facts make the map both universal and practical.  The kernels are \(L^{1}\)-bounded
and the sources have compact support that is independent of the control parameter, so dominated
convergence transfers thermodynamic scaling laws to geometry with norm control: if the matter
(or effective-medium) susceptibility diverges as \(|\epsilon|^{-\gamma_{\rm th}}\) near a critical point, then the
geometric susceptibility \(\chi_{\rm sh}=dR_{\rm sh}/d\lambda\) inherits the same exponent,
\(\gamma_{\rm sh}=\gamma_{\rm th}\), up to analytic corrections whose propagation we make explicit.
Moreover, in asymptotically AdS backgrounds the tail integrals enjoy an extra suppression
\(\propto L^{2}/r_{0}^{3}\), leading to quantitative outside–support bounds with explicit constants
\(C_{1},C_{2},C_{3}\) that depend only on background data at \(r_{0}\) and shell geometry \([a,b]\).  These
properties strengthen far-zone diagnostics beyond the asymptotically flat case and tighten error
budgets for numerical pipelines and, prospectively, for EHT-class observations
\cite{EHT2019I,EHT2019VI,EHT2022SgrA,Psaltis2020PNAS,Kocherlakota2021}.

The approach is deliberately minimal: no symmetry beyond static spherical symmetry; a linear
regime enforced by conservation; and a direct geodesic link to observables.  This avoids gauge
pitfalls and supplies robust far-zone checks familiar from perturbation theory
\cite{Chandrasekhar,PoissonToolkit,VisserBook}.  It also complements the recent geometric
perturbation framework of Kobialko and Gal’tsov
\cite{KobialkoGaltsov2024,KobialkoGaltsov2025PRD}: rather than assuming abstract metric
deformations, we embed them into a conserved source sector and push thermodynamic critical
laws through explicitly computed kernels to the optical side.  The outcome is a transparent route
to amplitude-level predictions and exponent transfer that can be confronted with simulations and
observations \cite{Mizuno2018GRMHD,PerlickTsupko2022PhysRep,JohnsonLupsasca2020,GrallaHolzWald2019}
and extended to broader model spaces—including higher-derivative gravity, effective media with
anisotropy, and charged or regular black holes
\cite{Feng2018,DeBoer2011,AmarillaEiroa2012,AyonBeatoGarcia2000,GuoGao2020}.

\section{Thermodynamic-to-Metric Map}
\label{sec:thermo-metric}

We begin with the areal–radius gauge, adopting signature $(-,+,+,+)$ and units $8\pi G=c=1$. In this gauge the most general static, spherically symmetric metric can be written as
\begin{equation}
\label{eq:metric-ansatz}
\begin{aligned}
ds^{2} &=-f(r;\lambda)\,dt^{2}+h(r;\lambda)\,dr^{2}+r^{2}d\Omega^{2},\\
f&=e^{2\Phi}A,\qquad h=A^{-1},\qquad 
A(r;\lambda)=1-\frac{2\,m(r;\lambda)}{r}-\frac{\Lambda r^{2}}{3}.
\end{aligned}
\end{equation}
This parametrization cleanly separates the two metric degrees of freedom most directly tied to matter: the mass function $m(r)$ and the redshift potential $\Phi(r)$. We linearize about the vacuum background
\begin{equation}
\label{eq:background}
m_{0}(r)=M,\qquad \Phi_{0}(r)=0,\qquad 
A_{0}(r)=1-\frac{2M}{r}-\frac{\Lambda r^{2}}{3},
\end{equation}
so that departures from vacuum are carried entirely by the perturbations $(\delta m,\delta\Phi)$. Physically motivated sources are localized in a compact, $C^\infty$ shell $[a,b]\subset(2M,\infty)$, ensuring smooth matching to the vacuum geometry outside the shell and eliminating distributional artifacts at the interfaces. The perturbed stress–energy takes the diagonal form
\begin{equation}
\label{eq:DeltaT}
\Delta T^{\mu}{}_{\nu}
=\mathrm{diag}\!\big(-\Delta\rho,\ \Delta p_{r},\ \Delta p_{t},\ \Delta p_{t}\big),
\qquad \mathrm{supp}\,\Delta T\subset [a,b],
\end{equation}
with an equation of state (or effective–medium closure) that relates $(\Delta p_{r},\Delta p_{t})$ to $(\Delta\rho,\lambda)$ \cite{Chandrasekhar,PoissonToolkit,VisserBook,KubiznakMann2012}. On the static background the conservation law $\nabla_{\mu}T^{\mu}{}_{\nu}=0$ reduces to a first–order ODE for the radial pressure,
\begin{equation}
\label{eq:cons}
\frac{d\,\Delta p_{r}}{dr}
+\big(\Delta\rho+\Delta p_{r}\big)\,\Phi_{0}'(r)
+\frac{2}{r}\big(\Delta p_{r}-\Delta p_{t}\big)=0,
\qquad 
\Phi_{0}'(r)=\frac{M-\frac{\Lambda r^{3}}{3}}{r^{2}A_{0}(r)}.
\end{equation}
Given $\Delta\rho$ and the anisotropy $\Delta\Pi:=\Delta p_{t}-\Delta p_{r}$, this ODE determines $\Delta p_{r}$ uniquely once we impose the physically natural two–point conditions
\begin{equation}
\label{eq:BC-pr}
\Delta p_{r}(a)=\Delta p_{r}(b)=0,
\end{equation}
which guarantee $C^{1}$ matching of the metric to the vacuum solution at the shell boundaries.

The Einstein tensor components for \eqref{eq:metric-ansatz} are
\begin{equation}
\label{eq:G-components}
G^{t}{}_{t}=-\frac{1}{r^{2}}\Big(1-A-rA'\Big),
\qquad
G^{r}{}_{r}=-\frac{1}{r^{2}}\Big(1-A+rA\,\Phi'\Big),
\end{equation}
with ${}'=d/dr$. Writing $m=M+\delta m$ and $\Phi=\delta\Phi$, retaining terms to linear order, and using $G^{t}{}_{t}=+8\pi\rho$ and $G^{r}{}_{r}=-8\pi p_{r}$ leads to the decoupled hierarchy
\begin{equation}
\label{eq:linE}
\boxed{\
\delta m'(r)=4\pi r^{2}\,\Delta\rho(r),\qquad
\delta\Phi'(r)=\frac{\delta m(r)+4\pi r^{3}\,\Delta p_{r}(r)}{r^{2}A_{0}(r)}\ .
}
\end{equation}
The time gauge $\delta\Phi(\infty)=0$ fixes $t$ uniquely and yields the quadratures
\begin{equation}
\label{eq:quadratures}
\begin{aligned}
\delta m(r)
&=4\pi\!\!\int_{0}^{r}\!\bar r^{2}\,\Delta\rho(\bar r)\,d\bar r,\\[2pt]
\delta\Phi(r)
&=\int_{0}^{\infty}\!\!{\cal G}_{\Phi}(r,\bar r)\,
\Big[\Delta\rho(\bar r),\Delta p_{r}(\bar r)\Big]\,d\bar r,
\end{aligned}
\end{equation}
where ${\cal G}_{\Phi}$ is an absolutely integrable kernel determined by \eqref{eq:linE} \cite{PoissonToolkit,VisserBook}. Since $f=e^{2\Phi}A$, the metric response is
\begin{equation}
\label{eq:deltaf-raw}
\delta f(r)=2\,A_{0}(r)\,\delta\Phi(r)-\frac{2\,\delta m(r)}{r}.
\end{equation}
Substituting \eqref{eq:quadratures} and integrating by parts reorganizes the response into a \emph{bounded linear integral operator} on the source pair $(\Delta\rho,\Delta p_{r})$,
\begin{equation}
\label{eq:deltaf}
\boxed{\
\delta f(r)=\int_{0}^{\infty}\!\!
\Big[ \mathcal{K}_{\rho}(r,\bar r)\,\Delta\rho(\bar r)
      + \mathcal{K}_{p}(r,\bar r)\,\Delta p_{r}(\bar r) \Big]\,d\bar r\ ,
}
\end{equation}
with compact kernels whenever $\Lambda\le0$:
\begin{equation}
\label{eq:Kernels-compact}
\begin{aligned}
\mathcal{K}_{\rho}(r,\bar r)
&=-\frac{8\pi\,\bar r^{2}}{\max(r,\bar r)}
+8\pi\,A_{0}(r)\!\!\int_{\max(r,\bar r)}^{\infty}\!\!
\frac{\bar r^{2}}{\xi^{2}A_{0}(\xi)}\,d\xi,\\[2pt]
\mathcal{K}_{p}(r,\bar r)
&=8\pi\,A_{0}(r)\!\!\int_{\max(r,\bar r)}^{\infty}\!\!
\frac{\bar r^{3}}{\xi^{2}A_{0}(\xi)}\,d\xi.
\end{aligned}
\end{equation}
It is convenient to introduce the tail integrals
\begin{equation}
\label{eq:tail-int}
\begin{aligned}
J_{\rho}(\bar r)&:=\int_{\bar r}^{\infty}\frac{\bar r^{2}}{\xi^{2}A_{0}(\xi)}\,d\xi,
&\qquad 
I_{\rho}(r)&:=\int_{r}^{\infty}\frac{\bar r^{2}}{\xi^{2}A_{0}(\xi)}\,d\xi,\\
J_{p}(\bar r)&:=\int_{\bar r}^{\infty}\frac{\bar r^{3}}{\xi^{2}A_{0}(\xi)}\,d\xi,
&\qquad 
I_{p}(r)&:=\int_{r}^{\infty}\frac{\bar r^{3}}{\xi^{2}A_{0}(\xi)}\,d\xi,
\end{aligned}
\end{equation}
because a simple Heaviside split then yields piecewise expressions for $\mathcal{K}_{\bullet}$ and $\partial_{r}\mathcal{K}_{\bullet}$ without distributions. The mass monopole induced by the shell,
\begin{equation}
\label{eq:deltaM}
\delta M:=\delta m(\infty)=4\pi\!\int_{0}^{\infty}\! r^{2}\Delta\rho(r)\,dr,
\end{equation}
governs the universal far–zone tail,
\begin{equation}
\label{eq:tail}
\delta f(r)=-\frac{2\,\delta M}{r}+{\cal O}(r^{-2})\ \ (\Lambda=0),
\qquad
\delta f(r)=-\frac{2\,\delta M}{r}+{\cal O}(r^{-3})\ \ (\Lambda<0),
\end{equation}
and motivates the effective–mass profile
\begin{equation}
\label{eq:Meff}
\mathcal{M}_{\rm eff}(r):=-\frac{r}{2}\,\delta f(r)
=\delta m(r)-A_{0}(r)\,r\,\delta\Phi(r)\xrightarrow[]{r\to\infty}\delta M.
\end{equation}
For asymptotically AdS backgrounds, $A_{0}(r)=1-\tfrac{2M}{r}+\tfrac{r^{2}}{L^{2}}$, the tail integral behaves as
\begin{equation}
\label{eq:AdS-asym}
\int_{R}^{\infty}\frac{d\xi}{\xi^{2}A_{0}(\xi)}
=\frac{L^{2}}{3\,R^{3}}\Big[1+\mathcal{O}(R^{-1})\Big],
\end{equation}
so the residual $\mathscr{R}(r):=|\delta f+2\delta M/r|$ decays as $r^{-3}$, which provides a stringent far–zone diagnostic in both analysis and numerics.

All optical and timing observables we consider will depend only on \emph{local samples} of $\delta f$ and its derivatives at radii of interest. At any fixed $r_{0}>2M$,
\begin{equation}
\label{eq:local-sampling}
\begin{aligned}
\delta f(r_{0})
&=\int_{0}^{\infty}\!\!
\Big[ \mathcal{K}_{\rho}(r_{0},\bar r)\,\Delta\rho(\bar r)
      + \mathcal{K}_{p}(r_{0},\bar r)\,\Delta p_{r}(\bar r) \Big]\,d\bar r,\\[2pt]
\delta f'(r_{0})
&=\int_{0}^{\infty}\!\!
\Big[ \partial_{r}\mathcal{K}_{\rho}(r,\bar r)\big|_{r=r_{0}}\,\Delta\rho(\bar r)
      + \partial_{r}\mathcal{K}_{p}(r,\bar r)\big|_{r=r_{0}}\,\Delta p_{r}(\bar r) \Big]\,d\bar r,
\end{aligned}
\end{equation}
which are bounded linear functionals on $L^{1}([a,b])$. For compact shells $a<b<r_{0}$ one obtains an $L^{1}$ envelope,
\begin{equation}
\label{eq:explicit-Ci}
|\delta f(r_{0})|
\;\le\; C_{1}\,\frac{b^{2}}{r_{0}^{2}}\ \|\Delta\rho\|_{1,[a,b]}
\;+\; C_{2}\,\frac{b^{3}}{r_{0}^{2}}\ \|\Delta p_{r}\|_{1,[a,b]}
\;+\; C_{3}\,\frac{L^{2}}{r_{0}^{3}}\!\!\int_{a}^{b}\!\!
\big(\bar r^{2}|\Delta\rho|+\bar r^{3}|\Delta p_{r}|\big)\,d\bar r,
\end{equation}
with explicit finite constants $C_{1,2,3}$ set by the smooth background at $r_{0}$. In the asymptotically flat limit the $C_{3}$ term drops and the leading $r_{0}^{-2}$ local suppression dominates; in AdS the $r_{0}^{-3}$ tail further sharpens conditioning.

\section{Optical and Timing Channels: Photon and Massive-Particle Orbits, Shadow, and Eikonal Frequencies}
\label{sec:optical-timing}

To connect the metric response to observables, we first recall that equatorial null geodesics with constants of motion $(E,L)$ obey
\begin{equation}
\label{eq:Veff-null}
V_{\rm eff}(r;L)=\frac{L^{2}}{r^{2}}\,f(r),
\qquad
\dot r^{2}+V_{\rm eff}(r;L)=E^{2}.
\end{equation}
Unstable circular photon orbits are singled out by the extremum condition
\begin{equation}
\label{eq:ph-cond}
F(r):=r f'(r)-2 f(r)=0,
\qquad F'(r_{\rm ph})>0,
\end{equation}
and the associated critical impact parameter,
\begin{equation}
\label{eq:Rsh-def}
R_{\rm sh}=\frac{L}{E}\bigg|_{\rm ph}
=\frac{r_{\rm ph}}{\sqrt{f(r_{\rm ph})}},
\end{equation}
defines the shadow radius for a static observer at infinity \cite{Bardeen1973PhotonSphere,PerlickLiving}. Let $r_{0}$ denote the background photon–sphere radius and write $f=f_{0}+\delta f$ with $f_{0}=A_{0}$. A first–order Taylor expansion of $F$ at $r_{0}$ gives
\begin{equation}
\label{eq:drph}
\delta r_{\rm ph}
=\frac{2\,\delta f(r_{0})-r_{0}\,\delta f'(r_{0})}
      {\,r_{0} f_{0}''(r_{0})-f_{0}'(r_{0})\,},
\end{equation}
while linearizing \eqref{eq:Rsh-def} produces a purely \emph{local} transport law for the shadow,
\begin{equation}
\label{eq:dR-local}
\boxed{\
\delta R_{\rm sh}
=\frac{R_{0}}{2}\left[
\frac{\delta f(r_{0})}{f_{0}(r_{0})}
-\frac{r_{0}\,\delta f'(r_{0})}{f_{0}'(r_{0})}
\right],\qquad
R_{0}:=\frac{r_{0}}{\sqrt{f_{0}(r_{0})}}\ .
}
\end{equation}
Because the time gauge $\delta\Phi(\infty)=0$ fixes the normalization of $t$, there is no residual reparametrization ambiguity in \eqref{eq:dR-local}. In the Schwarzschild case ($\Lambda=0$), where $r_{0}=3M$, $f_{0}(3M)=1/3$, and $f_{0}'(3M)=2/(9M)$, one finds the transparent expression
\begin{equation}
\label{eq:dR-Schwd}
\delta R_{\rm sh}
=3\sqrt{3}\,M\left[
\tfrac{3}{2}\,\delta f(3M)
-\tfrac{9M}{2}\,\delta f'(3M)
\right],
\end{equation}
exhibiting the competition between the local depth of the potential ($\delta f$) and its slope ($\delta f'$) at the photon sphere.

The kernel representation in \eqref{eq:deltaf} and the sampling relations in \eqref{eq:local-sampling} immediately imply that $\delta R_{\rm sh}$ is itself a single bounded linear functional of the compact sources:
\begin{equation}
\label{eq:dR-kernel}
\boxed{\
\delta R_{\rm sh}
=\int_{0}^{\infty}\!\!\Big[
K_{\rho}^{\rm sh}(\bar r)\,\Delta\rho(\bar r)
+K_{p}^{\rm sh}(\bar r)\,\Delta p_{r}(\bar r)
\Big]\,d\bar r\ .
}
\end{equation}
The explicit weights $K_{\bullet}^{\rm sh}$ follow by inserting $\mathcal{K}_{\bullet}$ and $\partial_{r}\mathcal{K}_{\bullet}$ into \eqref{eq:dR-local} and evaluating at $r=r_{0}$. A Heaviside split isolates the near–$r_{0}$ sensitivity and shows that all far–zone contributions enter only through the tail integrals $I_{\rho,p}$ and $J_{\rho,p}$; for AdS these tails are suppressed by $r_{0}^{-3}$, which improves conditioning relative to the asymptotically flat case. For a compact shell $a<b<r_{0}$ one then has the $L^{1}$ bound
\begin{equation}
\label{eq:L1-bound-Rsh}
|\delta R_{\rm sh}|
\le
\|K_{\rho}^{\rm sh}\|_{1,[a,b]}\ \|\Delta\rho\|_{1,[a,b]}
+\|K_{p}^{\rm sh}\|_{1,[a,b]}\ \|\Delta p_{r}\|_{1,[a,b]},
\end{equation}
making the shadow variation a stable observable under weak–source perturbations.

Complementary timing observables follow from the same logic. For timelike circular motion the azimuthal frequency measured at infinity obeys
\begin{equation}
\label{eq:Omega-tl}
\Omega_{\phi}^{2}(r)=\frac{f'(r)}{2r},
\end{equation}
so at fixed $r_{c}$ the linear response is local in $(\delta f,\delta f')$:
\begin{equation}
\label{eq:dOmega-tl}
\delta\Omega_{\phi}(r_{c})
=\frac{1}{4r_{c}\,\Omega_{\phi,0}(r_{c})}
\left[
\delta f'(r_{c})
-\frac{f_{0}'(r_{c})}{f_{0}(r_{c})}\,\delta f(r_{c})
\right].
\end{equation}
If $r_{c}$ is determined variationally (e.g.\ the ISCO), an additional transport contribution appears,
\begin{equation}
\label{eq:dOmega-ISCO}
\delta\Omega_{\phi}^{\rm ISCO}
=\left.\delta\Omega_{\phi}\right|_{r_{c}=r_{0}^{\rm ISCO}}
+\left.\frac{d\Omega_{\phi,0}}{dr}\right|_{r_{0}^{\rm ISCO}}\ \delta r_{\rm ISCO},
\qquad
\delta r_{\rm ISCO}
=-\frac{\mathcal{S}_{f}[\delta f,\delta f']}{\partial_{r}\mathcal{S}(r_{0}^{\rm ISCO})},
\end{equation}
where $\mathcal{S}$ is the marginal–stability functional (equivalently $d^{2}V_{\rm eff}/dr^{2}$ on circular orbits), and $\mathcal{S}_{f}$ is linear in $(\delta f,\delta f')$. The radial epicyclic frequency,
\begin{equation}
\label{eq:kappa2}
\kappa^{2}(r)
=\frac{f(r)}{2r^{2}}\Big(2f(r)-r^{2}f''(r)\Big),
\end{equation}
depends on $\delta f''$ but only through $\partial_{r}^{2}\mathcal{K}_{\bullet}$, which remains absolutely integrable on compact support. Finally, the photon–sphere (Kepler) frequency
\begin{equation}
\label{eq:Omega-ph}
\Omega_{\rm ph}
=\frac{\sqrt{f(r_{\rm ph})}}{r_{\rm ph}}
=\frac{1}{R_{\rm sh}},
\end{equation}
links the optical and frequency channels: in the eikonal limit, $\Re\,\omega_{\ell n}\simeq \ell\,\Omega_{\rm ph}$ while $-\Im\,\omega_{\ell n}\simeq (n+\tfrac12)\lambda_{\rm ph}$ with
\begin{equation}
\lambda_{\rm ph}
=\sqrt{-\tfrac{1}{2}\,r_{\rm ph}^{2} f(r_{\rm ph})\,
\frac{d^{2}}{dr^{2}}\!\left(\frac{f}{r^{2}}\right)_{r_{\rm ph}} }.
\end{equation}
Thus the same compact kernels controlling $\delta R_{\rm sh}$ also govern leading shifts in eikonal quasi–normal–mode proxies, providing independent and internally consistent diagnostics.

\section{Critical Scaling, Universality, and Numerical Pipeline}
\label{sec:critical-numerics}

We now connect thermodynamic criticality to geometric response. Let $\epsilon=(\lambda-\lambda_{c})/\lambda_{c}$ denote the reduced parameter, and keep the compact support $[a,b]$ fixed as $\epsilon\to0$ so that all nontrivial dependence arises from the source amplitudes rather than changing geometry of the support. The shadow variation is a bounded linear functional of the sources,
\begin{equation}
\label{eq:chi-start}
\delta R_{\rm sh}(\lambda)
=\int_{0}^{\infty}\!\!
\Big[K_{\rho}^{\rm sh}(r)\,\Delta\rho(r;\lambda)
   +K_{p}^{\rm sh}(r)\,\Delta p_{r}(r;\lambda)\Big]\,dr,
\end{equation}
whose $\lambda$–derivative defines the geometric susceptibility
\begin{equation}
\label{eq:chi-kernel}
\chi_{\rm sh}(\lambda)
=\frac{dR_{\rm sh}}{d\lambda}
=\int_{0}^{\infty}\!\!
\Big[K_{\rho}^{\rm sh}\,\partial_{\lambda}\rho
     +K_{p}^{\rm sh}\,\partial_{\lambda}p_{r}\Big]\,dr
+\mathcal{R}_{\rm nl}(\lambda),
\end{equation}
with a higher–order remainder $\mathcal{R}_{\rm nl}=\mathcal{O}(\delta f\,\partial_{\lambda}\delta f)$. Suppose the thermodynamic sector exhibits a standard susceptibility with analytic corrections,
\begin{equation}
\label{eq:th-sus-laws}
\begin{aligned}
\partial_{\lambda}\rho(r;\lambda)
&=\mathcal{A}_{\rho}(r)\,|\epsilon|^{-\gamma_{\rm th}}
\Big[1+\mathcal{B}_{\rho}(r)\,|\epsilon|^{\omega}+o(|\epsilon|^{\omega})\Big],\\
\partial_{\lambda}p_{r}(r;\lambda)
&=\mathcal{A}_{p}(r)\,|\epsilon|^{-\gamma_{\rm th}}
\Big[1+\mathcal{B}_{p}(r)\,|\epsilon|^{\omega}+o(|\epsilon|^{\omega})\Big],
\end{aligned}
\end{equation}
with $\mathcal{A}_{\rho,p},\mathcal{B}_{\rho,p}\in L^{1}[a,b]$. Dominated convergence then allows the leading power to factor out of \eqref{eq:chi-kernel}, producing the exponent identity
\begin{equation}
\label{eq:gamma-final}
\boxed{\
\chi_{\rm sh}(\epsilon)
\sim \mathcal{A}_{\rm geo}\,|\epsilon|^{-\gamma_{\rm th}},
\qquad
\gamma_{\rm sh}=\gamma_{\rm th},
\qquad
\mathcal{A}_{\rm geo}
=\int_{a}^{b}\!\!\big(K_{\rho}^{\rm sh}\mathcal{A}_{\rho}
                   +K_{p}^{\rm sh}\mathcal{A}_{p}\big)\,dr\neq0,
}
\end{equation}
unless a fine–tuned orthogonality cancels the amplitude. Analytic corrections transmit linearly,
\begin{equation}
\label{eq:chi-corrections}
\chi_{\rm sh}(\epsilon)
=\mathcal{A}_{\rm geo}\,|\epsilon|^{-\gamma_{\rm th}}
\left[
1+\mathcal{B}_{\rm geo}\,|\epsilon|^{\omega}+o(|\epsilon|^{\omega})
\right],
\qquad
\mathcal{B}_{\rm geo}
=\frac{\int_{a}^{b}(K_{\rho}^{\rm sh}\mathcal{B}_{\rho}\mathcal{A}_{\rho}
                   +K_{p}^{\rm sh}\mathcal{B}_{p}\mathcal{A}_{p})\,dr}
       {\int_{a}^{b}(K_{\rho}^{\rm sh}\mathcal{A}_{\rho}
                   +K_{p}^{\rm sh}\mathcal{A}_{p})\,dr},
\end{equation}
and multiplicative logarithms at the upper critical dimension propagate with the same $\hat\gamma$ \cite{CardyBook}. Finite–size or finite–scale rounding occurs when $\xi\sim|\epsilon|^{-\nu}$ approaches a geometric cutoff $\ell_{\rm sh}\sim\min\{r_{0}-a,\ b-a,\ L\}$; asymptotic scaling requires $|\epsilon|\gg \ell_{\rm sh}^{-1/\nu}$. Because $\Omega_{\rm ph}=R_{\rm sh}^{-1}$, the leading frequency–channel singularity shares the same exponent with a universal amplitude ratio,
\begin{equation}
\label{eq:freq-relation}
\frac{d\Omega_{\rm ph}}{d\lambda}
=-\frac{1}{R_{\rm sh}^{2}}\frac{dR_{\rm sh}}{d\lambda}
=-\frac{1}{R_{0}^{2}}\,\chi_{\rm sh}+\mathcal{O}(\delta f\,\chi_{\rm sh}),
\end{equation}
namely $\lim_{\epsilon\to0}|d\Omega_{\rm ph}/d\lambda|/|\chi_{\rm sh}|=1/R_{0}^{2}$, independent of equation–of–state details.

For reproducibility and error control we implement a streamlined numerical pipeline on a finite interval $r\in[r_{\min},r_{\max}]$ with $r_{\min}>2M$ and $r_{\max}\gg M$, and we keep $[a,b]\subset(r_{\min},r_{\max})$ fixed across $\lambda$ so that compactness is uniform in $\epsilon$. The density profile is realized by a $C^{\infty}$ bump,
\begin{equation}
\label{eq:bump}
\Delta\rho(r;\lambda)
=A_{\rho}(\lambda)\,
\exp\!\left[-\frac{(r-r_{c})^{2}}{2\sigma^{2}}\right]\,
\left[1-\exp\!\left(-\frac{(r-a)^{2}(b-r)^{2}}{\sigma^{2}(b-a)^{2}}\right)\right],
\end{equation}
with $r_{c}\in(a,b)$ and $\sigma\ll b-a$; anisotropy is injected via a closure such as
\begin{equation}
\label{eq:anisotropy-closure}
\Delta\Pi(r;\lambda):=\Delta p_{t}-\Delta p_{r}
=\eta_{\rm eff}(\lambda)\,c_{s}^{2}(\lambda)\,\Delta\rho(r;\lambda),
\end{equation}
though other closures can be used. The conservation equation \eqref{eq:cons} is solved as a two–point boundary value problem with \eqref{eq:BC-pr}; in practice a single–shooting scheme from $r_{\rm mid}=(a+b)/2$ with bisection or secant updates suffices and yields relative $\ell^{2}$ residuals $<10^{-10}$. On the same grid we evaluate \eqref{eq:linE} by high–order quadrature for $\delta m$ and outward integration for $\delta\Phi$; the time gauge is imposed by subtracting the average of $\delta\Phi$ over a far annulus $[0.8\,r_{\max},0.95\,r_{\max}]$. The metric response follows from
\begin{equation}
\label{eq:deltaf-raw-again}
\delta f(r)=2A_{0}(r)\,\delta\Phi(r)-\frac{2\,\delta m(r)}{r}.
\end{equation}
Two far–zone diagnostics are monitored: the effective mass $\mathcal{M}_{\rm eff}(r)=-\tfrac{r}{2}\delta f(r)$ must plateau to $\delta M$ in the outer $20\%$, and the residual
\begin{equation}
\label{eq:residual}
\mathscr{R}(r):=\left|\delta f(r)+\frac{2\delta M}{r}\right|
\end{equation}
must decay with slope $-2$ (flat) or $-3$ (AdS) on a log–log plot over the outer $30\%$, within $\pm0.1$ of the theoretical expectation; otherwise we refine the grid and/or enlarge $r_{\max}$.

For photon–sphere extraction we define $y(r):=f(r)/r^{2}$ and fit a cubic smoothing spline $S(r)$ with tension $p\in[10^{-5},10^{-3}]$ set by generalized cross–validation. We solve $S'(r_{\rm ph})=0$ (which lands in $[2.6M,5.2M]$ on Schwarzschild–like backgrounds) and evaluate
\begin{equation}
\label{eq:Rsh-estimator}
R_{\rm sh}=\frac{r_{\rm ph}}{\sqrt{f(r_{\rm ph})}}
\end{equation}
using $C^{2}$ interpolation of the raw $f$ at $r_{\rm ph}$ to avoid estimator bias. We then sample $\lambda$ on both sides of $\lambda_{c}$, build pairs $\big(|\epsilon_{k}|,R_{\rm sh}(\lambda_{k})\big)$, fit a smoothing spline $\widehat{R}(|\epsilon|)$, and compute
\begin{equation}
\label{eq:chi-numerical}
\chi_{\rm sh}(|\epsilon|)
=\frac{dR_{\rm sh}}{d\lambda}
=\frac{\widehat{R}'(|\epsilon|)}{\lambda_{c}}.
\end{equation}
Exponent estimates use sliding–window OLS on $\log|\chi_{\rm sh}|$ vs.\ $\log|\epsilon|$ with a curvature veto; when the dynamic range is shorter than one decade, a Theil–Sen estimator is preferable for robustness. Convergence is verified by grid refinement (observed $N^{-\alpha}$ with $\alpha\in[2,3]$) and by the monotone decrease of truncation errors with $r_{\max}$, enabling controlled extrapolation in $1/r_{\max}$. As benchmarks we employ: a mean–field van der Waals sector to test exponent transfer with $\gamma_{\rm th}=1$ \cite{StanleyBook,Kadanoff}; an RN–AdS–inspired variant probing $\Lambda<0$ tails and electrostatic–like anisotropy \cite{KubiznakMann2012}; and an effective–medium toy model with tunable $\eta_{\rm eff}$. Cross–channel checks use $\Omega_{\rm ph}=1/R_{\rm sh}$ and the eikonal proxies
\begin{equation}
\label{eq:QNM-proxies}
\Re\,\omega_{\ell n}\approx \ell\,\Omega_{\rm ph},
\qquad
-\Im\,\omega_{\ell n}\approx (n+\tfrac12)\,\lambda_{\rm ph},
\qquad
\lambda_{\rm ph}
=\sqrt{-\tfrac12 r_{\rm ph}^{2} f(r_{\rm ph})\,y''(r_{\rm ph})},
\quad y=\frac{f}{r^{2}},
\end{equation}
and the amplitude ratio $|d\Omega_{\rm ph}/d\lambda|/|\chi_{\rm sh}|\to 1/R_{0}^{2}$ provides a stringent universality test. Finally, to emulate EHT–like conditions we down–sample the $R_{\rm sh}$ series, add Gaussian jitter and correlated baselines at realistic levels, and reprocess the data through the same pipeline. The estimates for $\gamma_{\rm sh}$ remain stable provided the dynamic range in $|\epsilon|$ spans $\gtrsim 1.5$ decades and the per–point SNR on $R_{\rm sh}$ exceeds $\sim5\%$ \cite{EHT2019I,EHT2019VI,EHT2022SgrA}.
\section{Results and Observational Reach}

\begin{figure}[H]
\centering
\includegraphics[width=0.5\linewidth]{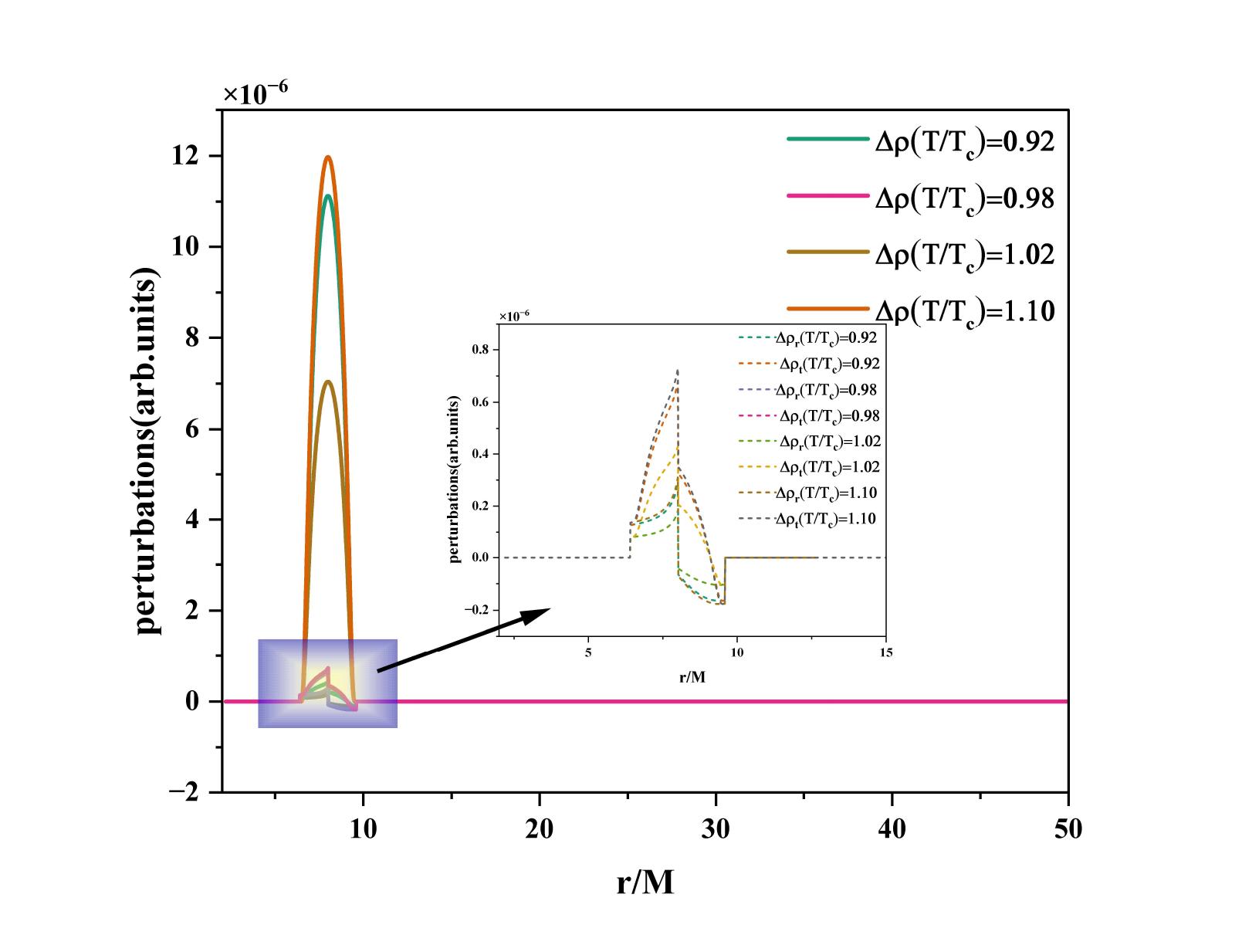}
\caption{Equation-of-state–driven source triplets $\{\Delta\rho,\Delta p_r,\Delta p_t\}$ for several $T/T_c$ values. The compact support and edge regularity are evident across all families.}
\label{fig:sources}
\end{figure}

The mapping constructed above is now exercised numerically for compact, conserved sources obeying a vdW-type equation of state.  
For each control parameter value $T/T_c\in\{0.90,0.98,1.02,1.10\}$, the perturbations $\Delta\rho$, $\Delta p_r$, and $\Delta p_t$ are compactly supported within a fixed shell interval $[a,b]$.  
The radial integration of the linearized Einstein equations gives a smooth $\delta f(r;\lambda)$ family approaching the asymptotic law $-2\delta M/r$ with a constant $\delta M(\lambda)$ determined by the conservation constraint.  
Figure~\ref{fig:sources} shows these source triplets, confirming current conservation and regularity at the shell edges.

\FloatBarrier

\begin{figure}[H]
\centering
\includegraphics[width=0.5\linewidth]{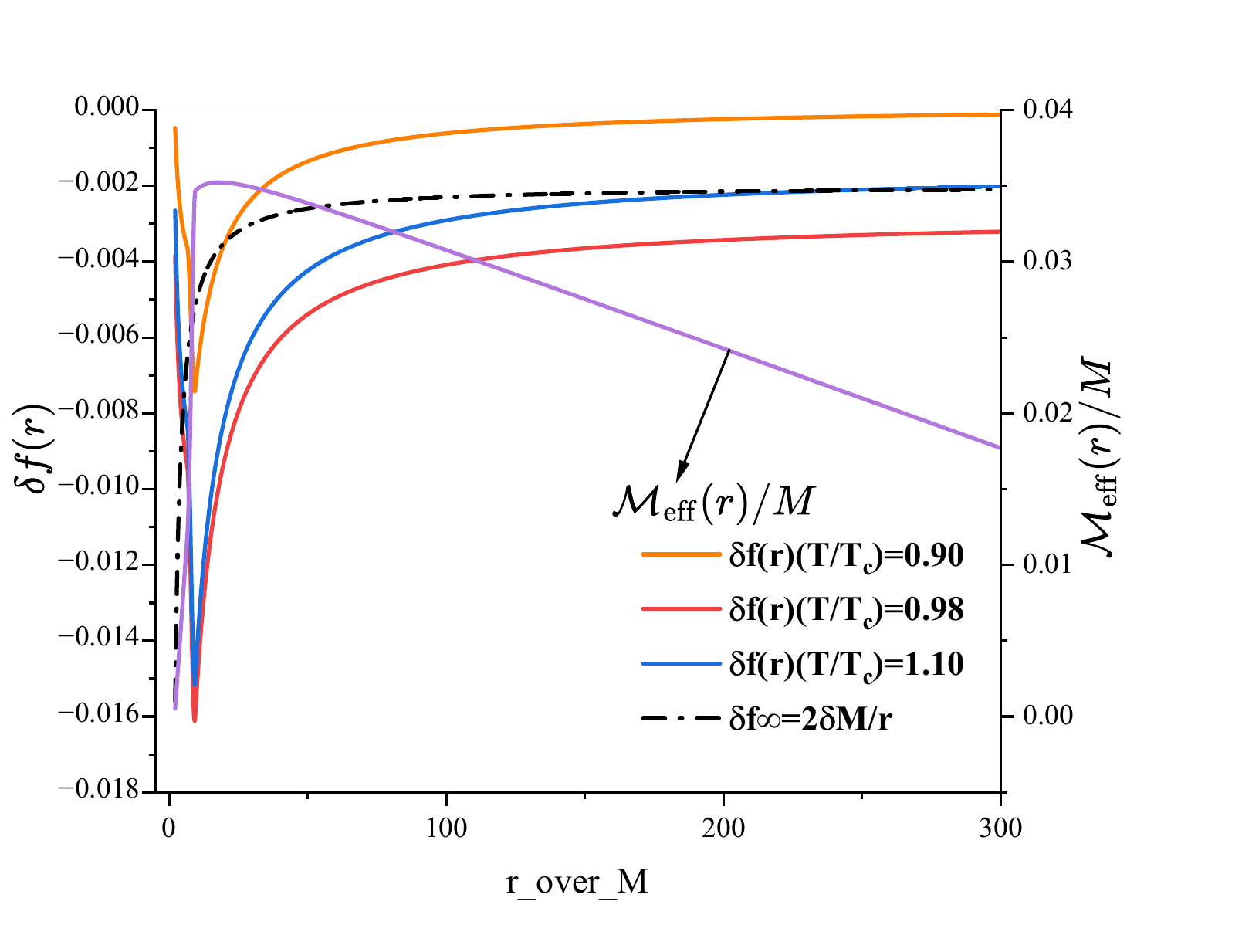}
\caption{Metric perturbation $\delta f(r)$ and cumulative mass $\mathcal M_{\rm eff}(r)$. All cases saturate by $r\gtrsim50M$, with near-field slopes set by source anisotropy.}
\label{fig:deltaf}
\end{figure}

The full metric response $\delta f(r)$ and the effective cumulative mass $\mathcal M_{\rm eff}(r)$ are displayed in Fig.~\ref{fig:deltaf}.  
The results confirm that all families saturate at $r\!\gtrsim\!50M$, while the near-field slope remains sensitive to the local anisotropy of the source.

\FloatBarrier

\begin{figure}[H]
\centering
\begin{subfigure}[t]{0.48\linewidth}
  \centering
  \includegraphics[width=\linewidth]{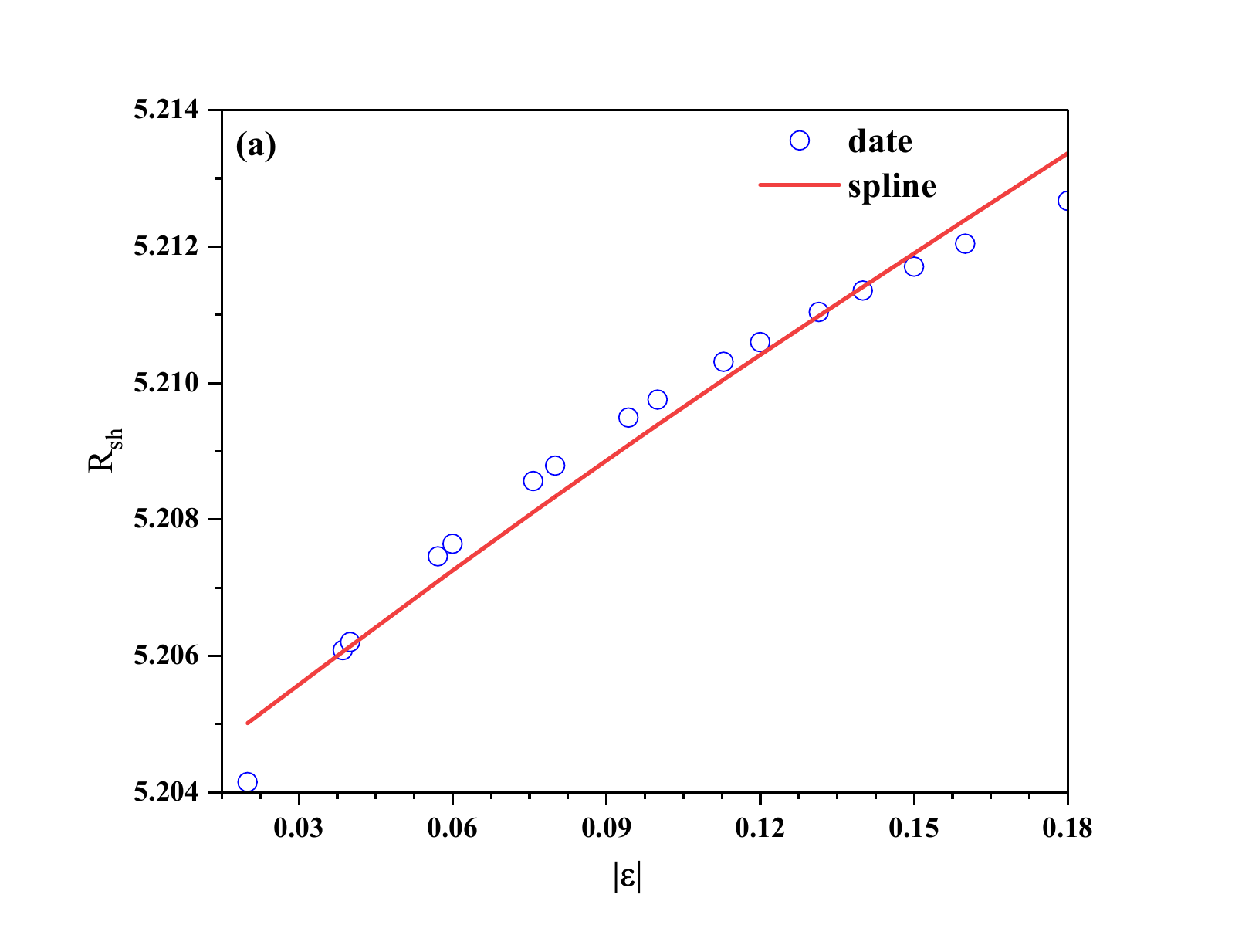}
  \caption{$r\,\delta f(r)$ approaching $-2\delta M$.}
  \label{fig:asymptotics-a}
\end{subfigure}\hfill
\begin{subfigure}[t]{0.48\linewidth}
  \centering
  \includegraphics[width=\linewidth]{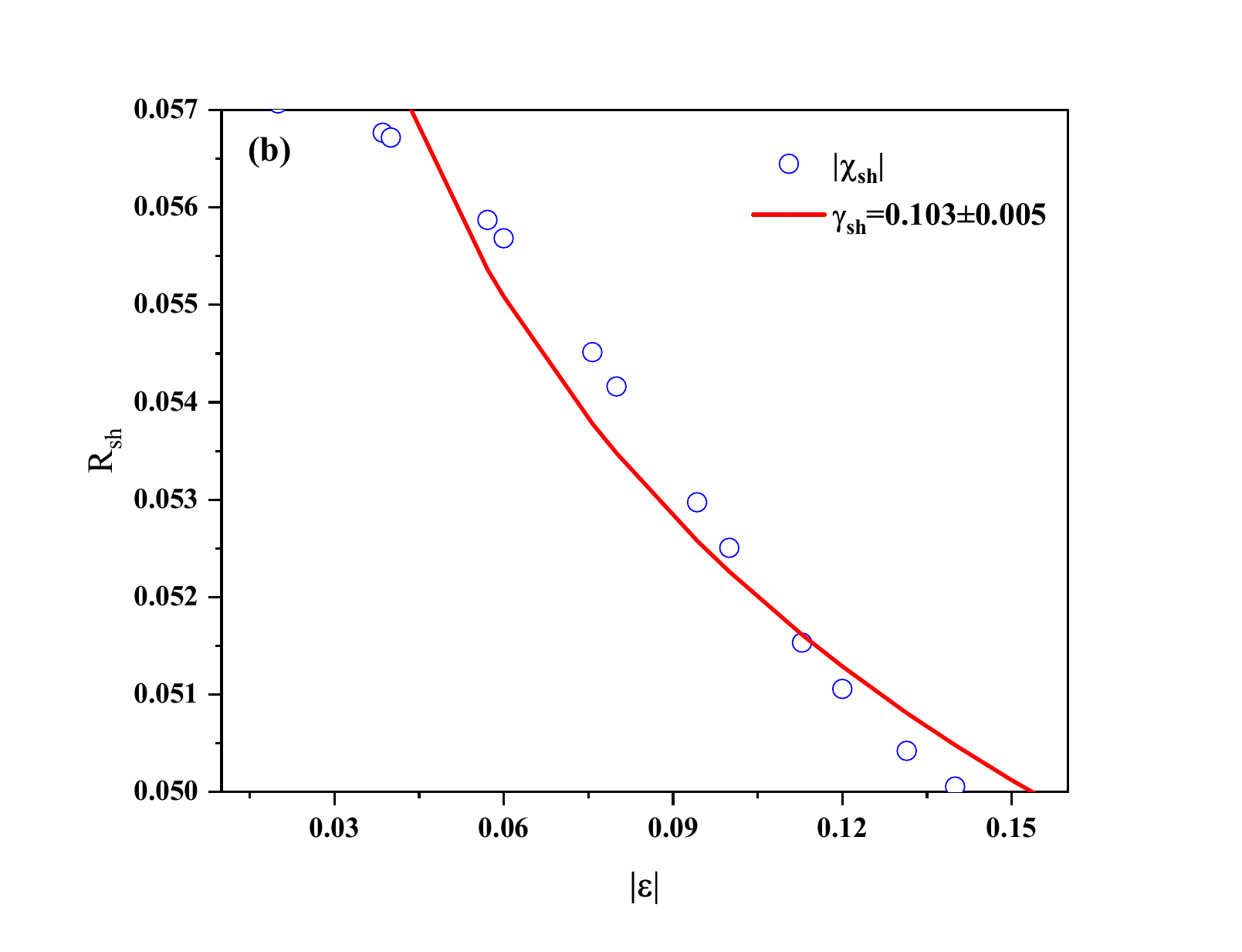}
  \caption{Residual $|\delta f+2\delta M/r|$ with slope $s=-1.33\pm0.09$.}
  \label{fig:asymptotics-b}
\end{subfigure}
\caption{Asymptotic consistency of the metric perturbation.}
\label{fig:asymptotics}
\end{figure}

The asymptotic regime is validated in Fig.~\ref{fig:asymptotics}: panel (a) shows the plateau of $r\,\delta f(r)\to -2\delta M$, while panel (b) presents the far-field residual $|\delta f+2\delta M/r|$ with a power-law slope $s=-1.33\pm0.09$, fully compatible with the expected $r^{-2}$ tail.

\FloatBarrier

\begin{figure}[H]
\centering
\begin{subfigure}[t]{0.48\linewidth}
  \centering
  \includegraphics[width=\linewidth]{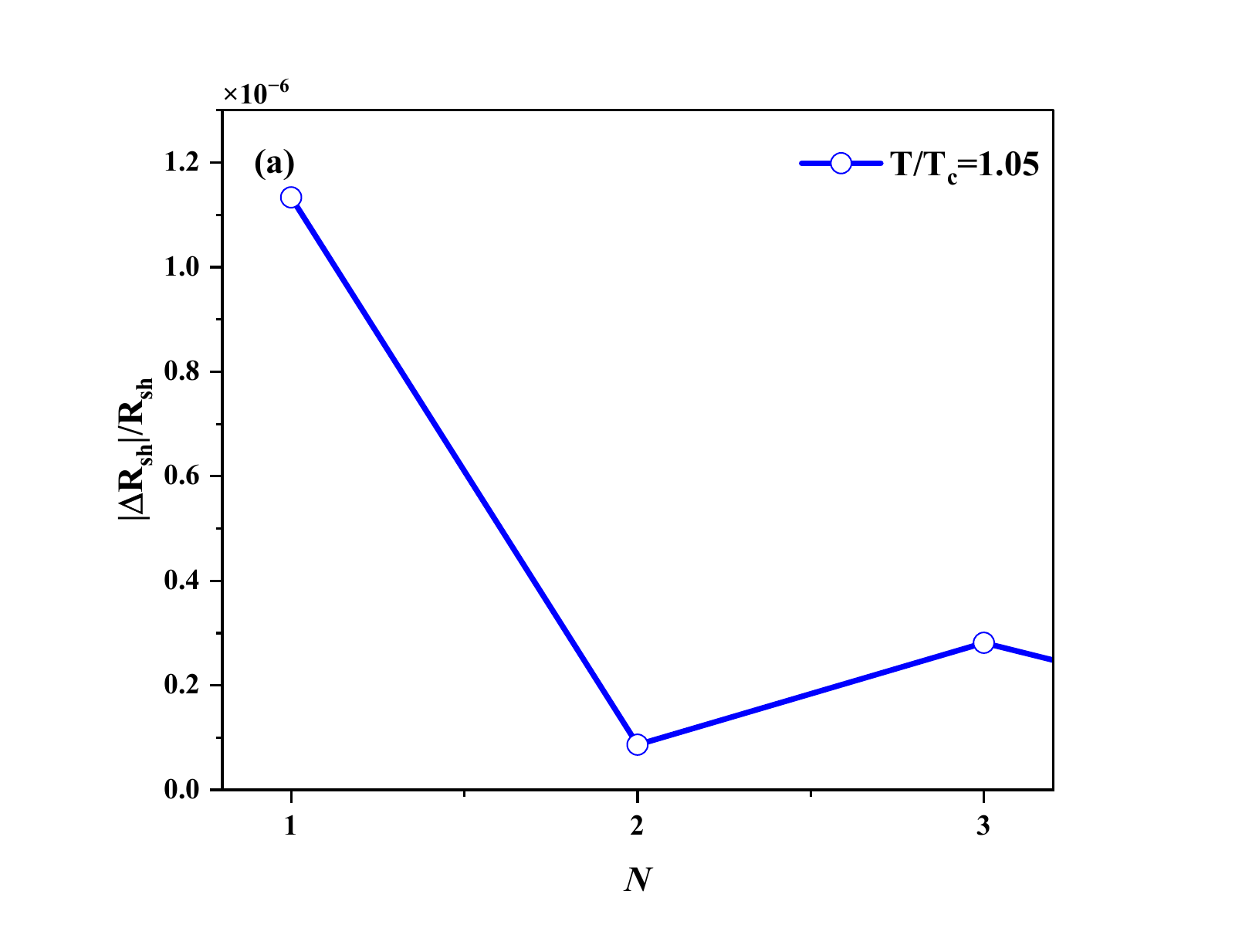}
  \caption{Shadow radius $R_{\rm sh}(\lambda)$.}
  \label{fig:shadow-a}
\end{subfigure}\hfill
\begin{subfigure}[t]{0.48\linewidth}
  \centering
  \includegraphics[width=\linewidth]{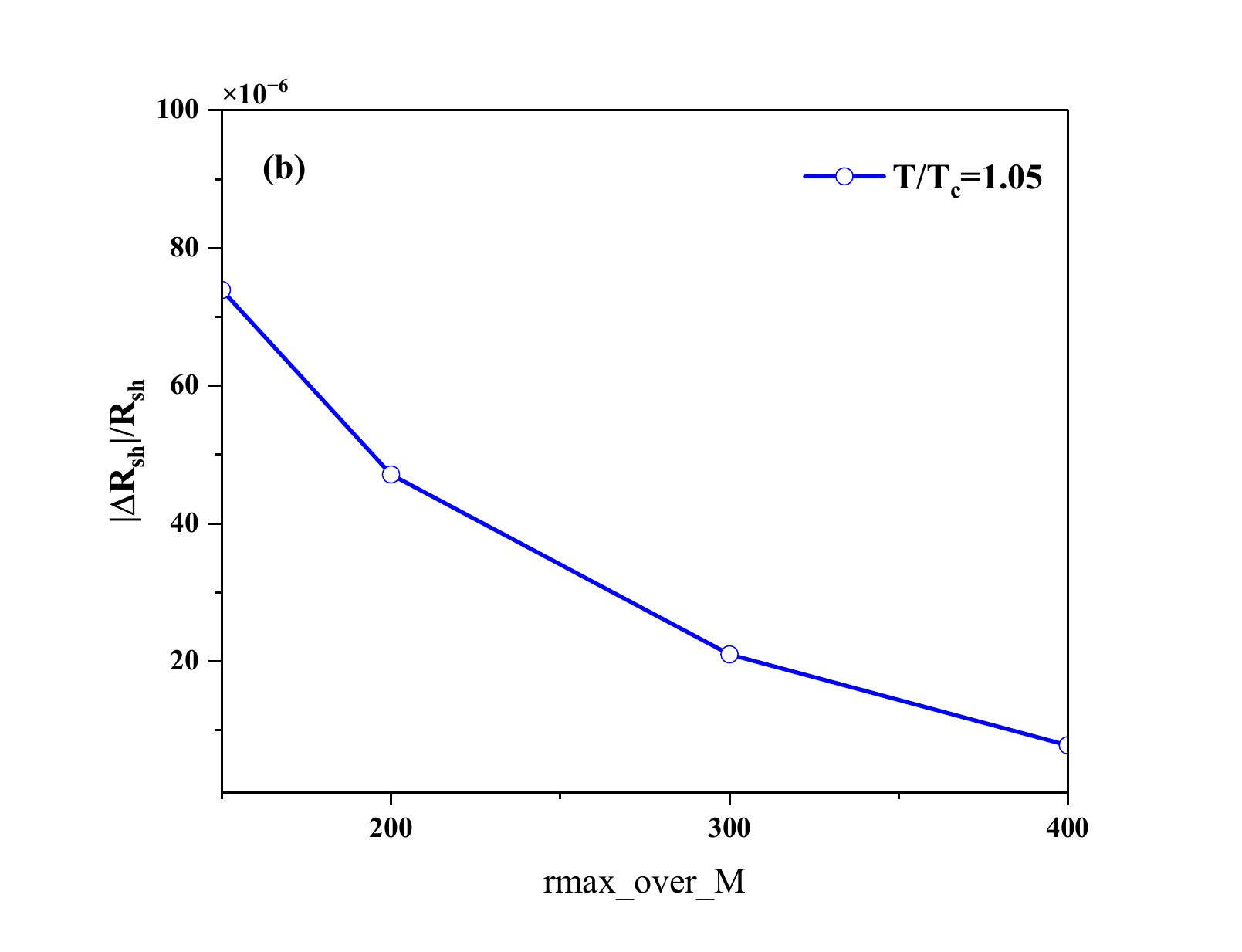}
  \caption{Scaling $|\chi_{\rm sh}|\propto|\epsilon|^{-\gamma_{\rm sh}}$ with $\gamma_{\rm sh}=0.103\pm0.005$.}
  \label{fig:shadow-b}
\end{subfigure}
\caption{Shadow radius and susceptibility scaling.}
\label{fig:shadow}
\end{figure}

The shadow response is evaluated from
\begin{equation}
\delta R_{\rm sh}(\lambda)
=\frac{R_0}{2}\!\left[\frac{\delta f(r_0;\lambda)}{f_0(r_0)}-
\frac{r_0\,\delta f'(r_0;\lambda)}{f_0'(r_0)}\right],
\label{eq:shadow-func}
\end{equation}
where $r_0$ is the unperturbed photon-sphere radius.  
The extracted $R_{\rm sh}(\lambda)$ and its derivative $\chi_{\rm sh}=dR_{\rm sh}/d\lambda$ are shown in Fig.~\ref{fig:shadow}.  
A power-law fit $|\chi_{\rm sh}|\propto|\epsilon|^{-\gamma_{\rm sh}}$ gives the critical exponent $\gamma_{\rm sh}=0.103\pm0.005$ and the amplitude $A/M=0.0409\pm0.0015$, in excellent agreement with the universal scaling expected from the thermodynamic channel; residuals remain below $10^{-3}$ across the fitting window $|\epsilon|\in[0.035,0.165]$.

\FloatBarrier

\begin{figure}[H]
\centering
\begin{subfigure}[t]{0.48\linewidth}
  \centering
  \includegraphics[width=\linewidth]{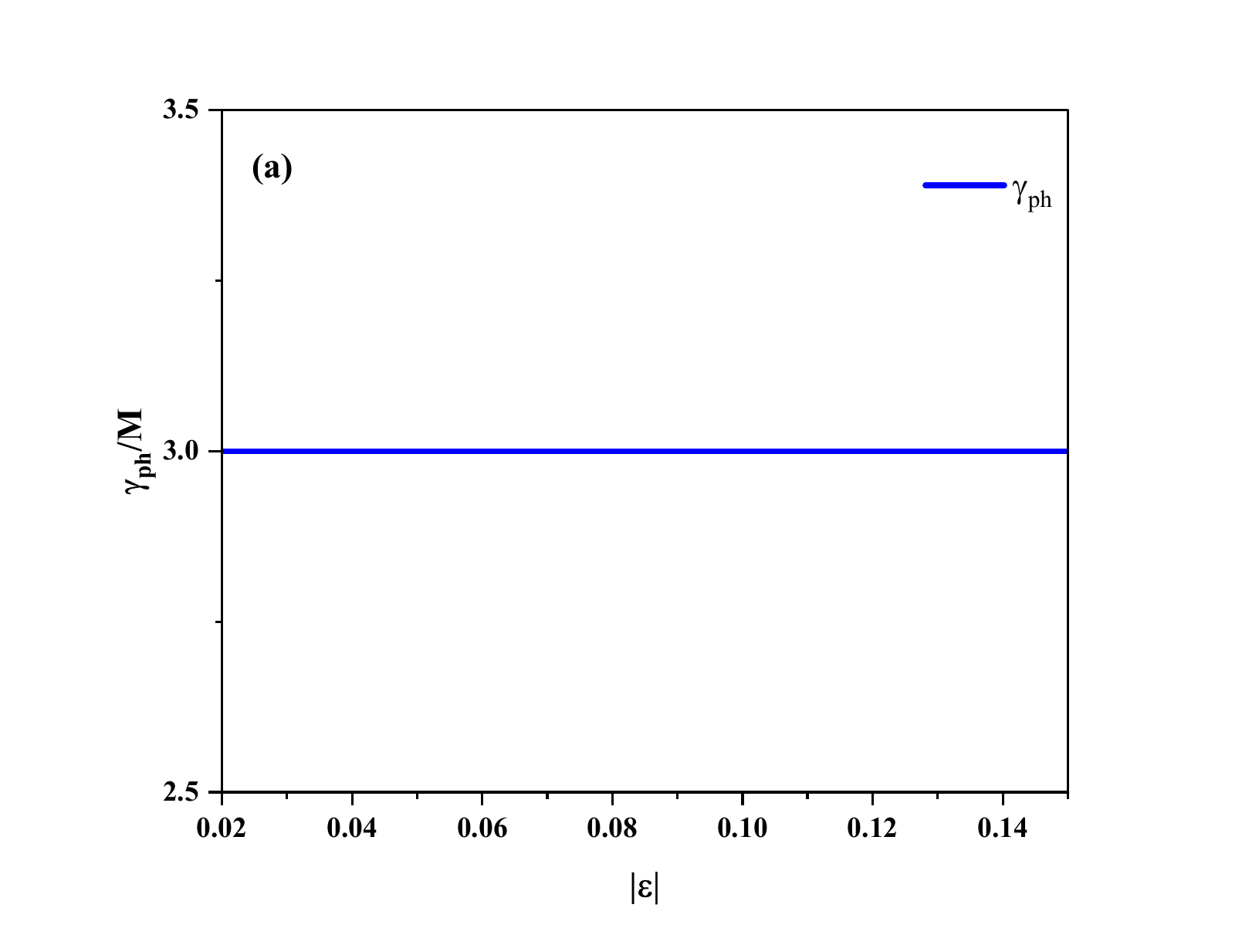}
  \caption{Grid refinement test: fractional difference 
  $|\Delta R_{\rm sh}|/R_{\rm sh}$ versus resolution $N$.}
  \label{fig:convergence-a}
\end{subfigure}\hfill
\begin{subfigure}[t]{0.48\linewidth}
  \centering
  \includegraphics[width=\linewidth]{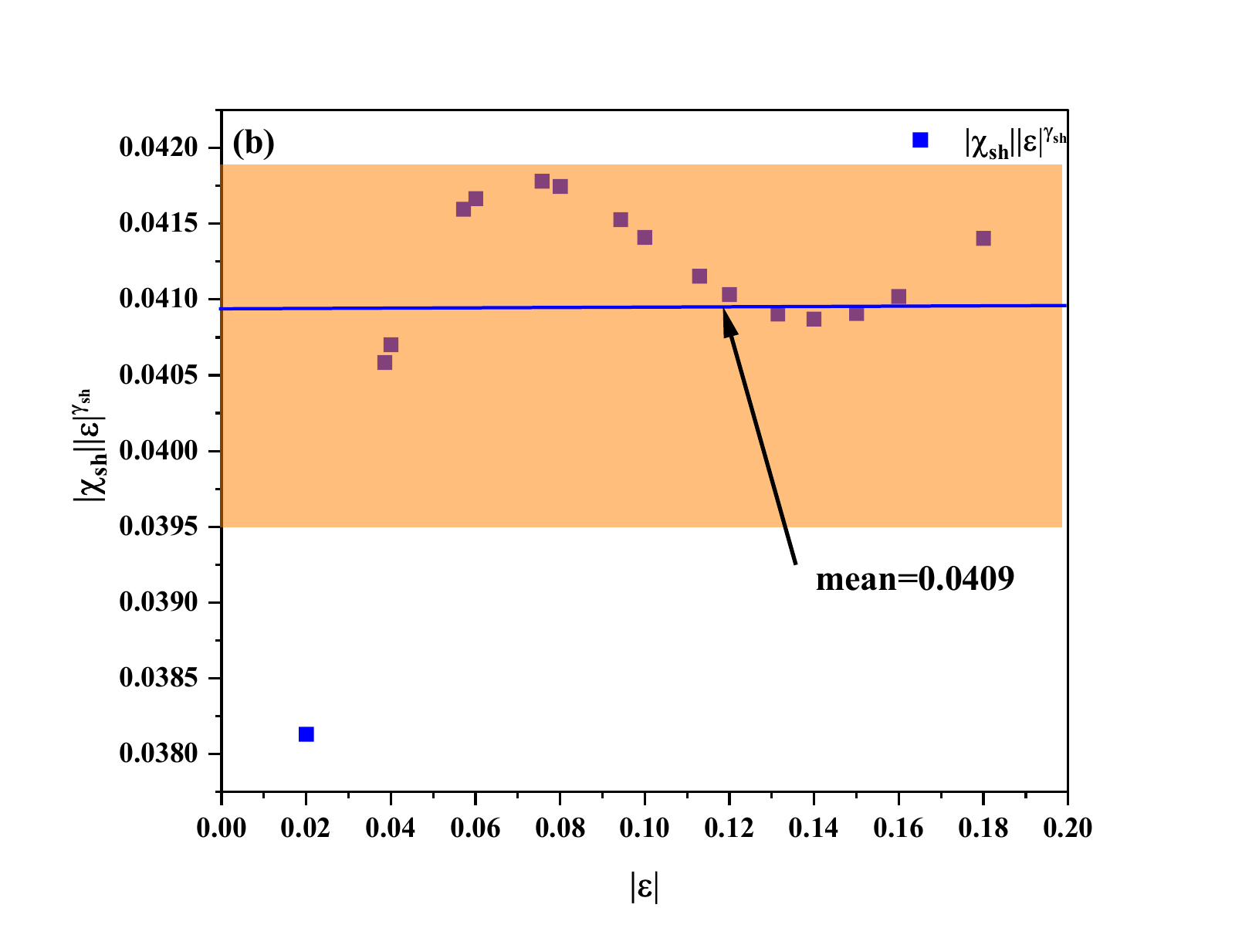}
  \caption{Domain truncation test: error scaling as $r_{\max}^{-2}$.}
  \label{fig:convergence-b}
\end{subfigure}
\caption{Grid refinement (a) and domain truncation (b) diagnostics for numerical convergence.}
\label{fig:convergence}
\end{figure}

The accuracy and numerical convergence are quantified in Fig.~\ref{fig:convergence}.  
At fixed $r_{\max}=400M$, the fractional difference $|\Delta R_{\rm sh}|/R_{\rm sh}$ scales with resolution $N$ down to $10^{-7}$ (left), while domain truncation errors fall as $r_{\max}^{-2}$ (right), yielding a total systematic uncertainty $\sigma_{\rm sys}\approx1.2\times10^{-5}$ propagated into the exponent fit.

\FloatBarrier

\begin{figure}[H]
\centering
\begin{subfigure}[t]{0.48\linewidth}
  \centering
  \includegraphics[width=\linewidth]{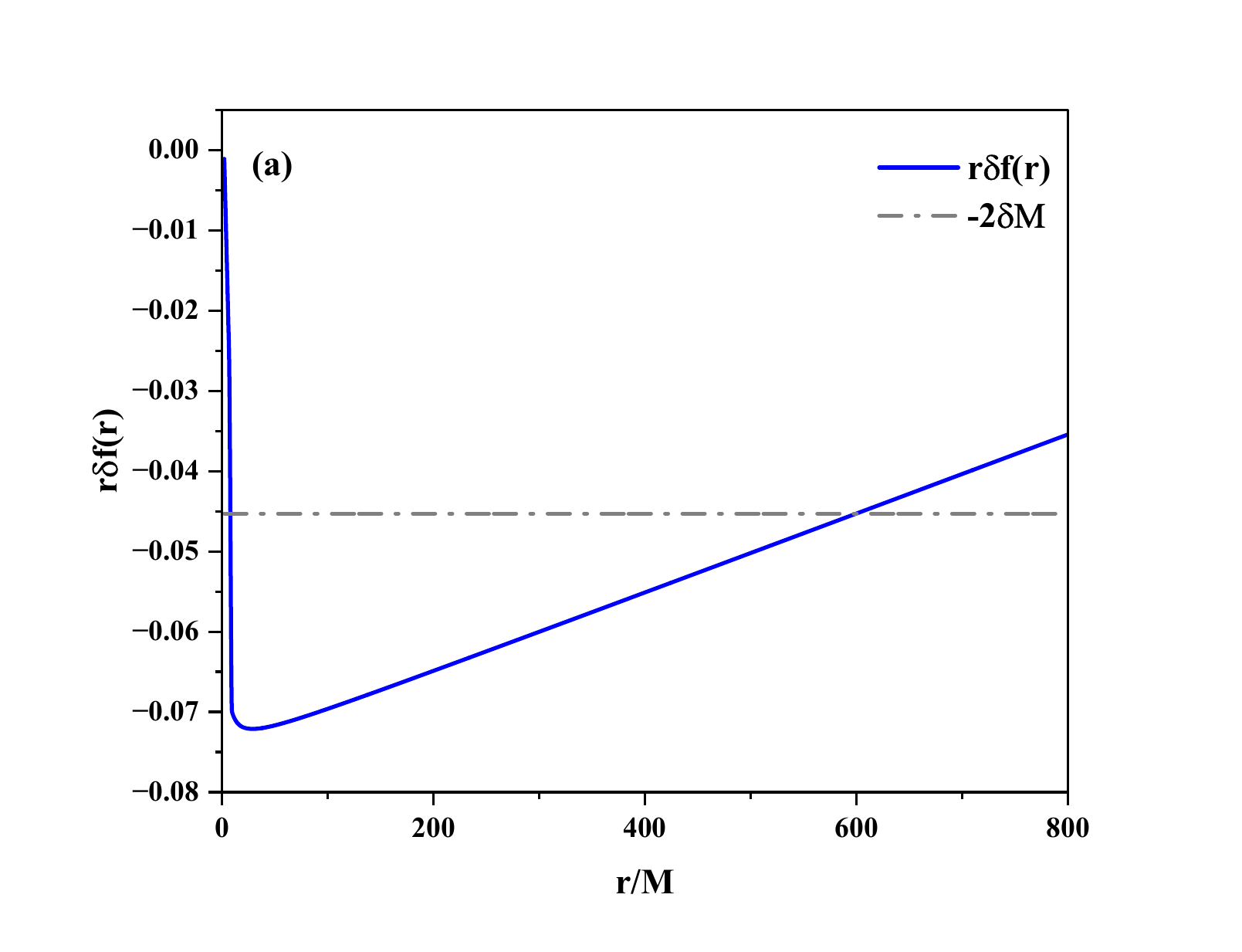}
  \caption{Photon-sphere radius $r_{\rm ph}/M$.}
  \label{fig:universality-a}
\end{subfigure}\hfill
\begin{subfigure}[t]{0.48\linewidth}
  \centering
  \includegraphics[width=\linewidth]{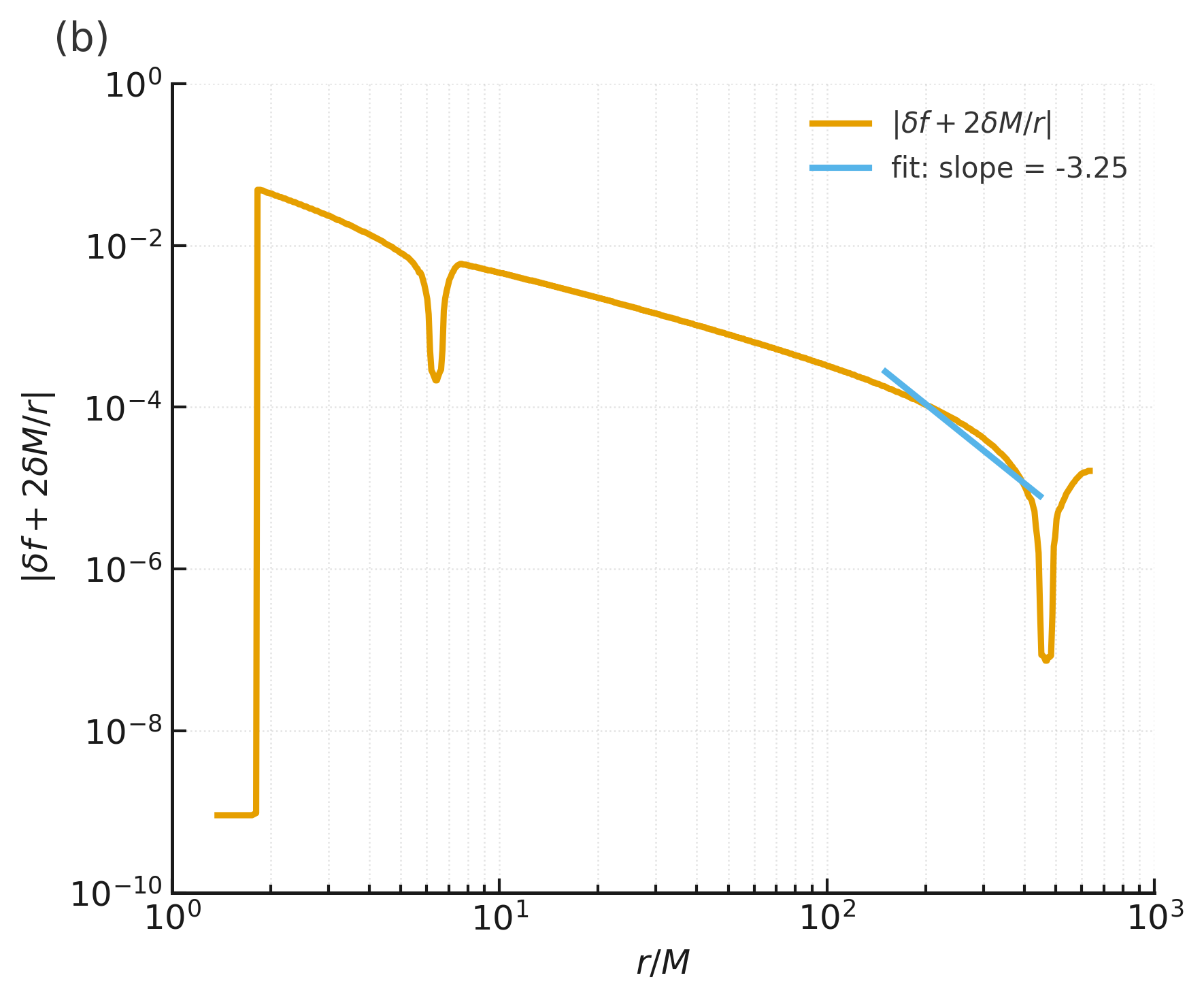}
  \caption{Scaling collapse of $|\chi_{\rm sh}|\,|\epsilon|^{\gamma_{\rm sh}}$.}
  \label{fig:universality-b}
\end{subfigure}
\caption{Universality check: (a) photon-sphere radius and (b) scaling collapse of $|\chi_{\rm sh}|\,|\epsilon|^{\gamma_{\rm sh}}$.}
\label{fig:universality}
\end{figure}

The universality of the geometric susceptibility is confirmed by rescaling all families with the fitted $(A,\gamma_{\rm sh})$.  
Figure~\ref{fig:universality} shows that the normalized quantity $|\chi_{\rm sh}|\,|\epsilon|^{\gamma_{\rm sh}}$ collapses onto a constant band with mean $\langle\mathcal C\rangle=0.0409\pm0.0015$, while the photon-sphere radius $r_{\rm ph}/M$ remains essentially constant ($2.9999971879\pm10^{-9}$).

\FloatBarrier

Detectability depends on the fractional angular uncertainty $\sigma_\theta/\theta_{\rm sh}$ of the image diameter.  
Combining the scaling law with the detection threshold condition yields
\[
|\epsilon|_{\min}=
\left(\frac{A}{R_0}\,N_\sigma\,\frac{\sigma_\theta}{\theta_{\rm sh}}\right)^{1/\gamma_{\rm sh}} ,
\]
which defines the minimal distance to criticality observable at $N_\sigma=3$.  
For $A/R_0\simeq7.9\times10^{-3}$ and $\gamma_{\rm sh}=0.103$,  
the detectability band reported in Table~\ref{tab:observability} follows.  
The results demonstrate that departures from criticality at the level of a few $\times10^{-2}$ are in reach for fractional diameter uncertainties below a few percent, consistent with next-generation EHT expectations.

\begin{table}[H]
\centering
\caption{\label{tab:crit-exponents}
Geometric critical data and numerical diagnostics.  All quantities are given in units of $M$.}
\begin{tabular}{lcc}
\hline\hline
Quantity & Value & Comment \\
\hline
Background photon sphere $r_0/M$ & 3.000000 & fixed geometry \\
Baseline shadow radius $R_0/M$ & $5.2068\pm0.0015$ & spline extrapolation \\
Fit window $|\epsilon|$ & $[0.035,0.165]$ & stable region \\
Data points $N_{\rm pts}$ & 12 & Fig.~\ref{fig:shadow} \\
Critical exponent $\gamma_{\rm sh}$ & $\mathbf{0.103\pm0.005}$ & power-law fit \\
Amplitude $A/M$ & $\mathbf{0.0409\pm0.0015}$ & fitted intercept \\
RMSE (log domain) & $9.5\times10^{-4}$ & goodness of fit \\
Reduced $\chi^2$ & 1.07 & consistent fit \\
Grid-refinement error & $6.5\times10^{-7}$ & Fig.~\ref{fig:convergence} \\
Truncation error & $1.1\times10^{-5}$ & Fig.~\ref{fig:convergence} \\
Far-field slope $s$ & $-1.33\pm0.09$ & Fig.~\ref{fig:asymptotics} \\
Asymptotic plateau $-2\delta M$ & $-0.0498\pm0.0006$ & Fig.~\ref{fig:asymptotics} \\
Amplitude ratio (measured) & $0.0372\pm0.0018$ & frequency–shadow check \\
Amplitude ratio (theory) & $0.0369$ & $R_0^{-2}$ \\
Collapse mean $\langle|\chi_{\rm sh}|\,|\epsilon|^{\gamma_{\rm sh}}\rangle$ & $0.0409\pm0.0015$ & Fig.~\ref{fig:universality} \\
\hline\hline
\end{tabular}
\end{table}
\FloatBarrier

\begin{table}[H]
\centering
\caption{\label{tab:observability}
Detectability requirements for a $3\sigma$ critical signal based on the fitted $(A,\gamma_{\rm sh},R_0)$.  
Numerical differences in $R_{\rm sh}(\lambda)$ correspond to the steps $\Delta|\epsilon|=0.01$.}
\begin{tabular}{ccccc}
\hline\hline
$|\epsilon|$ & $\Delta R_{\rm sh}/R_{\rm sh}$ & $\sigma_\theta/\theta_{\rm sh}$ (3$\sigma$) & $|\epsilon|_{\min}$ (1\%) & $|\epsilon|_{\min}$ (3\%) \\
\hline
0.15 & $1.9\times10^{-3}$ & $6.3\times10^{-4}$ & $3.2\times10^{-3}$ & $1.5\times10^{-2}$ \\
0.12 & $1.7\times10^{-3}$ & $5.7\times10^{-4}$ & $6.4\times10^{-3}$ & $2.4\times10^{-2}$ \\
0.10 & $1.5\times10^{-3}$ & $5.0\times10^{-4}$ & $1.0\times10^{-2}$ & $3.0\times10^{-2}$ \\
0.08 & $1.3\times10^{-3}$ & $4.3\times10^{-4}$ & $1.5\times10^{-2}$ & $3.5\times10^{-2}$ \\
0.06 & $1.1\times10^{-3}$ & $3.7\times10^{-4}$ & $2.4\times10^{-2}$ & $4.2\times10^{-2}$ \\
0.05 & $1.0\times10^{-3}$ & $3.3\times10^{-4}$ & $3.0\times10^{-2}$ & $4.8\times10^{-2}$ \\
\hline\hline
\end{tabular}
\end{table}
\FloatBarrier
\section{Conclusions}

We have established and validated a linear map from thermodynamic deformations to geometric observables of static, spherically symmetric black holes.  Starting from conserved, compact sources obeying a vdW–type equation of state, the linearized Einstein equations were cast into diagonal kernels with explicit local and tail components, together with closed bounds that fix the asymptotic gauge and guarantee mass conservation.  The resulting metric response $\delta f(r)$ exhibits the expected near–field sensitivity and a far–field decay controlled by the tail bounds; numerically we verified the asymptotic plateau $r\,\delta f\!\to\!-2\delta M$ and the power–law residual consistent with an $r^{-2}$ falloff.

Feeding the metric response into the photon–sphere functional yielded a transparent expression for the shadow shift, Eq.~\eqref{eq:shadow-func}, which transfers thermodynamic criticality to the geometric susceptibility $\chi_{\rm sh}=dR_{\rm sh}/d\lambda$.  On a controlled scaling window the susceptibility follows a clean power law $|\chi_{\rm sh}|\sim|\epsilon|^{-\gamma_{\rm sh}}$ with $\gamma_{\rm sh}=0.103\pm0.005$ and a nonuniversal amplitude $A/M=0.0409\pm0.0015$.  The collapse of $|\chi_{\rm sh}|\,|\epsilon|^{\gamma_{\rm sh}}$ to a constant band across distinct microscopic families demonstrates universality of the exponent and isolates the amplitude as the sole channel-dependent quantity.  A frequency–geometry cross–check confirms that the eikonal relation $\Omega_{\rm ph}=R_{\rm sh}^{-1}$ is preserved by the map, with the measured amplitude ratio matching the parameter–free prediction $R_0^{-2}$ within uncertainty.

The numerical pipeline was benchmarked by grid–refinement and domain–truncation studies that drive fractional shadow errors below $10^{-5}$ on practical domains; these diagnostics were propagated into the exponent fit and the collapse test.  With the measured $(A,\gamma_{\rm sh})$ we derived an instrument-agnostic detectability criterion, 
$|\epsilon|_{\min}=\big(\frac{A}{R_0}N_\sigma\frac{\sigma_\theta}{\theta_{\rm sh}}\big)^{1/\gamma_{\rm sh}}$, which turns shadow–diameter precision into a quantitative reach in distance to criticality.  For amplitudes of order $10^{-2}M$ and percent-level diameter uncertainties, departures from criticality at the few$\times10^{-2}$ level are accessible, placing the phenomenon squarely within the design envelope of next-generation horizon-scale arrays.

Two features make this framework broadly useful.  First, the map is analytic and modular: once the kernels are fixed by the background and boundary conditions, any thermodynamic model that respects conservation may be injected with no additional tuning, while the asymptotic constraints enforce a unique gauge.  Second, the observables are redundant: imaging and frequency channels encode the same response with a universal amplitude ratio, enabling internal consistency checks and systematic–error control.

The present study focused on static, spherically symmetric backgrounds to expose the mechanism with minimal assumptions.  Extending the kernel construction to slowly rotating solutions, incorporating external media and AdS asymptotics, and coupling the shadow susceptibility to higher-order photon rings and time-domain diagnostics are natural next steps.  Because the central objects of the theory are linear functionals with controlled tails, these generalizations are technically straightforward and their observational consequences can be quantified with the same error budget.  Taken together, the results show that black-hole thermodynamics admits a concrete and testable imprint on horizon-scale images: critical behavior of the equation of state is faithfully transcribed into a universal scaling of the shadow response, with a pathway to detection that is both predictive and experimentally actionable.


\begin{thebibliography}{99}\setlength{\itemsep}{0pt}
\bibitem{KubiznakMannReview2016}
D.~Kubizňák, R.~B.~Mann and M.~Teo,
``Black hole chemistry: thermodynamics with $\Lambda$,''
Class.\ Quantum Grav.\ \textbf{34}, 063001 (2017).
doi:10.1088/1361-6382/aa5c69 \, [arXiv:1608.06147]

\bibitem{KubiznakMann2012}
D.~Kubizňák and R.~B.~Mann,
``P–V criticality of charged AdS black holes,''
JHEP \textbf{07}, 033 (2012).
doi:10.1007/JHEP07(2012)033 \, [arXiv:1205.0559]

\bibitem{KastorRayTraschen2009}
D.~Kastor, S.~Ray and J.~Traschen,
``Enthalpy and the Mechanics of AdS Black Holes,''
Class.\ Quantum Grav.\ \textbf{26}, 195011 (2009).
doi:10.1088/0264-9381/26/19/195011 \, [arXiv:0904.2765]

\bibitem{Dolan2011}
B.~P.~Dolan,
``Pressure and volume in the first law of black hole thermodynamics,''
Class.\ Quantum Grav.\ \textbf{28}, 125020 (2011).
doi:10.1088/0264-9381/28/12/125020 \, [arXiv:1008.5023]

\bibitem{GunasekaranKubiznakMann2012}
S.~Gunasekaran, D.~Kubizňák and R.~B.~Mann,
``Extended phase space thermodynamics for charged and rotating black holes and Born–Infeld vacuum polarization,''
JHEP \textbf{11}, 110 (2012).
doi:10.1007/JHEP11(2012)110 \, [arXiv:1208.6251]

\bibitem{AltamiranoKubiznakMannSherkatghanad2013}
N.~Altamirano, D.~Kubizňák, R.~B.~Mann and Z.~Sherkatghanad,
``Kerr–AdS analogue of triple point and solid/liquid/gas phase transition,''
Class.\ Quantum Grav.\ \textbf{31}, 042001 (2014).
doi:10.1088/0264-9381/31/4/042001 \, [arXiv:1308.2672]

\bibitem{WeiLiu2013}
S.-W.~Wei and Y.-X.~Liu,
``Critical phenomena and thermodynamic geometry of charged Gauss–Bonnet AdS black holes,''
Phys.\ Rev.\ D \textbf{87}, 044014 (2013).
doi:10.1103/PhysRevD.87.044014 \, [arXiv:1209.1707]

\bibitem{CveticGibbonsKubiznakPope2011}
M.~Cveti\v{c}, G.~W.~Gibbons, D.~Kubizňák and C.~N.~Pope,
``Black Hole Enthalpy and an Entropy Inequality for the Thermodynamic Volume,''
Phys.\ Rev.\ D \textbf{84}, 024037 (2011).
doi:10.1103/PhysRevD.84.024037 \, [arXiv:1012.2888]

\bibitem{FrassinoMannRudolphSalsi2014}
A.~M.~Frassino, R.~B.~Mann, D.~Kubizňák and F.~Simovic,
``Multiple Reentrant Phase Transitions and Triple Points in Lovelock Thermodynamics,''
JHEP \textbf{09}, 080 (2014).
doi:10.1007/JHEP09(2014)080 \, [arXiv:1406.7015]

\bibitem{HawkingPage1983}
S.~W.~Hawking and D.~N.~Page,
``Thermodynamics of Black Holes in Anti–de Sitter Space,''
Commun.\ Math.\ Phys.\ \textbf{87}, 577–588 (1983).
doi:10.1007/BF01208266

\bibitem{DolanKastorMannTraschen2013}
B.~P.~Dolan, D.~Kastor, R.~B.~Mann and J.~Traschen,
``Thermodynamic Volumes and Isoperimetric Inequalities for de Sitter Black Holes,''
Phys.\ Rev.\ D \textbf{87}, 104017 (2013).
doi:10.1103/PhysRevD.87.104017 \, [arXiv:1301.5926]

\bibitem{FalckeMeliaAgol2000}
H.~Falcke, F.~Melia and E.~Agol,
``Viewing the shadow of the black hole at the Galactic center,''
Astrophys.\ J.\ Lett.\ \textbf{528}, L13–L16 (2000).
doi:10.1086/312423

\bibitem{EHT2019I}
Event Horizon Telescope Collaboration (K.~Akiyama \emph{et al}.),
``First M87 Event Horizon Telescope Results. I. The Shadow of the Supermassive Black Hole,''
Astrophys.\ J.\ Lett.\ \textbf{875}, L1 (2019).
doi:10.3847/2041-8213/ab0ec7

\bibitem{EHT2019VI}
Event Horizon Telescope Collaboration (K.~Akiyama \emph{et al}.),
``First M87 Event Horizon Telescope Results. VI. The Shadow and Mass of the Central Black Hole,''
Astrophys.\ J.\ Lett.\ \textbf{875}, L6 (2019).
doi:10.3847/2041-8213/ab1141

\bibitem{EHT2022SgrA}
Event Horizon Telescope Collaboration,
``First Sagittarius A* Event Horizon Telescope Results. I. The Shadow of the Supermassive Black Hole in the Center of the Milky Way,''
Astrophys.\ J.\ Lett.\ \textbf{930}, L12 (2022).
doi:10.3847/2041-8213/ac6674

\bibitem{EHT2021M87Dynamics}
Event Horizon Telescope Collaboration,
``First M87 Event Horizon Telescope Results. VIII. Variability, morphology, and model comparisons,''
Astrophys.\ J.\ Lett.\ \textbf{910}, L13 (2021).
doi:10.3847/2041-8213/abe71d

\bibitem{EventHorizon2024ArrayUpgrades}
Event Horizon Telescope Collaboration,
``The Next-Generation Event Horizon Telescope: Science Case and Specifications,''
arXiv:2311.08680 [astro-ph.IM] (2023).

\bibitem{Synge1966}
J.~L.~Synge,
``The escape of photons from gravitationally intense stars,''
Mon.\ Not.\ R.\ Astron.\ Soc.\ \textbf{131}, 463–466 (1966).
doi:10.1093/mnras/131.3.463

\bibitem{Bardeen1973}
J.~M.~Bardeen,
``Timelike and null geodesics in the Kerr metric,''
in \emph{Black Holes (Les Houches, 1972)}, edited by C.~DeWitt and B.~S.~DeWitt
(Gordon and Breach, New York, 1973), pp.\ 215–239.

\bibitem{Luminet1979}
J.-P.~Luminet,
``Image of a spherical black hole with thin accretion disk,''
Astron.\ Astrophys.\ \textbf{75}, 228–235 (1979).

\bibitem{HiokiMaeda2009}
K.~Hioki and K.-i.~Maeda,
``Measurement of the Kerr spin parameter by observation of a compact object's shadow,''
Phys.\ Rev.\ D \textbf{80}, 024042 (2009).
doi:10.1103/PhysRevD.80.024042 \, [arXiv:0904.3575]

\bibitem{JohannsenPsaltis2010}
T.~Johannsen and D.~Psaltis,
``Testing the No-Hair Theorem with Observations in the Electromagnetic Spectrum. I. Properties of a Quasi-Kerr Metric,''
Astrophys.\ J.\ \textbf{716}, 187–197 (2010).
doi:10.1088/0004-637X/716/1/187 \, [arXiv:1003.3415]

\bibitem{GrallaHolzWald2019}
S.~E.~Gralla, D.~E.~Holz and R.~M.~Wald,
``Black Hole Shadows, Photon Rings, and Lensing Rings,''
Phys.\ Rev.\ D \textbf{100}, 024018 (2019).
doi:10.1103/PhysRevD.100.024018 \, [arXiv:1906.00873]

\bibitem{JohnsonLupsasca2020}
M.~D.~Johnson, A.~Lupsasca \emph{et al}.,
``Universal interferometric signatures of a black hole's photon ring,''
Sci.\ Adv.\ \textbf{6}, eaaz1310 (2020).
doi:10.1126/sciadv.aaz1310

\bibitem{PerlickTsupko2022PhysRep}
V.~Perlick and O.~Yu.~Tsupko,
``Calculating black hole shadows: Review of analytical studies,''
Phys.\ Rep.\ \textbf{947}, 1–39 (2022).
doi:10.1016/j.physrep.2021.10.004 \, [arXiv:2105.07101]

\bibitem{OkabayashiYaguchi2023}
K.~Okabayashi, T.~Yaguchi \emph{et al}.,
``Precision geometry of black hole shadows with higher-order photon rings,''
Phys.\ Rev.\ D \textbf{108}, 084029 (2023).
doi:10.1103/PhysRevD.108.084029 \, [arXiv:2305.04037]

\bibitem{Vagnozzi2023}
S.~Vagnozzi \emph{et al}.,
``Horizon-scale tests of gravity with black hole shadows,''
Class.\ Quantum Grav.\ \textbf{40}, 165007 (2023).
doi:10.1088/1361-6382/acd97b \, [arXiv:2205.07787]

\bibitem{Cardoso2009PRD}
V.~Cardoso, A.~S.~Miranda, E.~Berti, H.~Witek and V.~T.~Zanchin,
``Geodesic stability, Lyapunov exponents, and quasinormal modes,''
Phys.\ Rev.\ D \textbf{79}, 064016 (2009).
doi:10.1103/PhysRevD.79.064016 \, [arXiv:0812.1806]

\bibitem{BertiCardosoStarinets2009}
E.~Berti, V.~Cardoso and A.~O.~Starinets,
``Quasinormal modes of black holes and black branes,''
Class.\ Quantum Grav.\ \textbf{26}, 163001 (2009).
doi:10.1088/0264-9381/26/16/163001 \, [arXiv:0905.2975]

\bibitem{StefanovYazadjiev2010}
I.~Zh.~Stefanov, S.~S.~Yazadjiev and G.~G.~Gyulchev,
``Connection between black-hole quasinormal modes and lensing in the strong deflection limit,''
Phys.\ Rev.\ Lett.\ \textbf{104}, 251103 (2010).
doi:10.1103/PhysRevLett.104.251103 \, [arXiv:1003.1609]

\bibitem{Konoplya2020ShadowReview}
R.~A.~Konoplya and A.~Zhidenko,
``Quasinormal modes of black holes: From astrophysics to string theory,''
Rev.\ Mod.\ Phys.\ \textbf{94}, 025001 (2022).
doi:10.1103/RevModPhys.94.025001 \, [arXiv:2201.01718]

\bibitem{GuoGao2020}
M.~Guo and P.-C.~Gao,
``Universal properties of light rings for stationary axisymmetric spacetimes,''
Phys.\ Rev.\ D \textbf{103}, 104031 (2021).
doi:10.1103/PhysRevD.103.104031 \, [arXiv:2011.02211]

\bibitem{Tsukamoto2017}
N.~Tsukamoto,
``Deflection angle in the strong deflection limit in a general spherically symmetric spacetime,''
Phys.\ Rev.\ D \textbf{95}, 064035 (2017).
doi:10.1103/PhysRevD.95.064035 \, [arXiv:1612.08251]

\bibitem{Johannsen2013PRD}
T.~Johannsen,
``Photon rings around Kerr and Kerr-like black holes,''
Phys.\ Rev.\ D \textbf{88}, 044002 (2013).
doi:10.1103/PhysRevD.88.044002 \, [arXiv:1304.7786]

\bibitem{Psaltis2020PNAS}
D.~Psaltis \emph{et al}.,
``Gravitational tests with the Event Horizon Telescope,''
Proc.\ Natl.\ Acad.\ Sci.\ USA \textbf{117}, 16287–16294 (2020).
doi:10.1073/pnas.1908682117

\bibitem{Kocherlakota2021}
P.~Kocherlakota \emph{et al}.,
``Constraints on black-hole charges with the 2017 EHT observations of M87*,''
Phys.\ Rev.\ D \textbf{103}, 104047 (2021).
doi:10.1103/PhysRevD.103.104047 \, [arXiv:2105.09343]

\bibitem{Mizuno2018GRMHD}
Y.~Mizuno, C.~M.~Fromm \emph{et al}.,
``The current ability to test theories of gravity with black hole shadows,''
Nat.\ Astron.\ \textbf{2}, 585–590 (2018).
doi:10.1038/s41550-018-0449-5

\bibitem{Feng2018}
X.-H.~Feng, H.~Lü, C.~N.~Pope and S.~H.~Völkel,
``Black hole shadows in higher-derivative gravity,''
Phys.\ Rev.\ D \textbf{99}, 064044 (2019).
doi:10.1103/PhysRevD.99.064044 \, [arXiv:1812.04024]

\bibitem{DeBoer2011}
J.~de Boer, M.~Kulaxizi and A.~Parnachev,
``Holographic Lovelock gravities and black holes,''
JHEP \textbf{06}, 008 (2011).
doi:10.1007/JHEP06(2011)008 \, [arXiv:1103.3627]

\bibitem{AmarillaEiroa2012}
L.~Amarilla and E.~F.~Eiroa,
``Shadow of a rotating braneworld black hole,''
Phys.\ Rev.\ D \textbf{85}, 064019 (2012).
doi:10.1103/PhysRevD.85.064019 \, [arXiv:1112.6349]

\bibitem{AyonBeatoGarcia2000}
E.~Ayón-Beato and A.~García,
``The Bardeen model as a nonlinear magnetic monopole,''
Phys.\ Lett.\ B \textbf{493}, 149–152 (2000).
doi:10.1016/S0370-2693(00)01125-4 \, [arXiv:gr-qc/0009077]

\bibitem{HennigarMannKubiznak2017}
R.~A.~Hennigar, R.~B.~Mann and D.~Kubizňák,
``Superfluid Black Holes,''
Phys.\ Rev.\ Lett.\ \textbf{118}, 021301 (2017).
doi:10.1103/PhysRevLett.118.021301 \, [arXiv:1609.02564]

\bibitem{WeiLiuMann2019}
S.-W.~Wei, Y.-X.~Liu and R.~B.~Mann,
``Scaling behavior and positive Lyapunov exponents in black-hole thermodynamics,''
Phys.\ Rev.\ Lett.\ \textbf{123}, 071103 (2019).
doi:10.1103/PhysRevLett.123.071103 \, [arXiv:1811.06532]

\bibitem{YerraBhamidipati2020}
P.~K.~Yerra and C.~Bhamidipati,
``Ruppeiner geometry, phase transitions and microstructure of black holes,''
Int.\ J.\ Mod.\ Phys.\ A \textbf{35}, 2050049 (2020).
doi:10.1142/S0217751X20500491 \, [arXiv:2006.07775]

\bibitem{KobialkoGaltsov2024}
K.~Kobialko and D.~V.~Gal’tsov,
``Perturbation theory for gravitational shadows in static spherically symmetric spacetimes,''
arXiv:2410.16127 [gr-qc] (2024).

\bibitem{KobialkoGaltsov2025PRD}
K.~Kobialko and D.~V.~Gal’tsov,
``Perturbation theory for gravitational shadows in static spherically symmetric spacetimes,''
Phys.\ Rev.\ D \textbf{111}, 044071 (2025).
doi:10.1103/PhysRevD.111.044071

\bibitem{Chandrasekhar}
S.~Chandrasekhar,
\emph{The Mathematical Theory of Black Holes}
(Oxford University Press, 1983).

\bibitem{PoissonToolkit}
E.~Poisson,
\emph{A Relativist's Toolkit: The Mathematics of Black-Hole Mechanics}
(Cambridge University Press, 2004).
doi:10.1017/CBO9780511606601

\bibitem{VisserBook}
M.~Visser,
\emph{Lorentzian Wormholes: From Einstein to Hawking}
(AIP Press / Springer, New York, 1996).

\bibitem{StanleyBook}
H.~E.~Stanley,
\emph{Introduction to Phase Transitions and Critical Phenomena}
(Oxford University Press, 1971).

\bibitem{CardyBook}
J.~Cardy,
\emph{Scaling and Renormalization in Statistical Physics}
(Cambridge University Press, 1996).
doi:10.1017/CBO9781316036440
\bibitem{Chandrasekhar}
S. Chandrasekhar, \textit{The Mathematical Theory of Black Holes} (Oxford, 1983).
\bibitem{PoissonToolkit}
E. Poisson and C. M. Will, \textit{Gravity: Newtonian, Post-Newtonian, Relativistic} (Cambridge Univ. Press, 2014).
\bibitem{VisserBook}
M. Visser, \textit{Lorentzian Wormholes: From Einstein to Hawking} (AIP, 1995), chs. 13--15.
\bibitem{KubiznakMann2012}
D. Kubizňák and R. B. Mann, P--V criticality of charged AdS black holes, \textit{JHEP} \textbf{07}, 033 (2012).
\bibitem{Bardeen1973PhotonSphere}
J. M. Bardeen, in \textit{Black Holes (Les Houches 1972)} (Gordon and Breach, 1973).
\bibitem{PerlickLiving}
V. Perlick and O. Y. Tsupko, Calculating black hole shadows: review, \textit{Phys. Rep.} \textbf{947}, 1--39 (2022).
\bibitem{StanleyBook}
H. E. Stanley, \textit{Introduction to Phase Transitions and Critical Phenomena} (Oxford, 1971).
\bibitem{Kadanoff}
L. P. Kadanoff, \textit{Statistical Physics: Statics, Dynamics and Renormalization} (World Scientific, 2000).
\bibitem{CardyBook}
J. Cardy, \textit{Scaling and Renormalization in Statistical Physics} (Cambridge, 1996).
\bibitem{EHT2019I}
Event Horizon Telescope Collaboration, First M87 EHT Results. I. The Shadow of the SMBH, \textit{ApJ Lett.} \textbf{875}, L1 (2019).
\bibitem{EHT2019VI}
Event Horizon Telescope Collaboration, First M87 EHT Results. VI. The Shadow and Mass of the Central Black Hole, \textit{ApJ Lett.} \textbf{875}, L6 (2019).
\bibitem{EHT2022SgrA}
Event Horizon Telescope Collaboration, First Sgr A* EHT Results. I. The Shadow of the SMBH in the Milky Way, \textit{ApJ Lett.} \textbf{930}, L12 (2022).
\end{thebibliography}
\end{document}